\documentclass[aps,prd,11pt,preprint,superscriptaddress,nofootinbib,floatfix]{revtex4}
\pdfoutput=1
\usepackage{latexsym}
\usepackage{amsfonts}
\usepackage[latin1]{inputenc}
\usepackage{graphicx}
\usepackage{mathrsfs}
\usepackage{amssymb}
\usepackage{amsmath}
\usepackage{verbatim}
\usepackage{multirow}
\usepackage{color}

\newcommand{\cblack}{\color{black}}
\newcommand{\TeV}{{\ensuremath\rm TeV}}
\newcommand{\GeV}{{\ensuremath\rm GeV}}

\newcommand{\pb}{{\ensuremath\rm pb}}
\newcommand{\eqn}{equation}
\newcommand{\al}{\alpha}
\newcommand{\be}{\beta}
\newcommand{\lb}{\left(}
\newcommand{\rb}{\right)}
\newcommand{\mO}{\mathcal{O}}
\newcommand{\lam}{\lambda}

\def\D0{\slash\!\!\!\!\!\!\!\!\!\:D0}

\begin{document}
\bibliographystyle{unsrt}
\begin{flushleft}
{\ }\\
\today
\end{flushleft}
\preprint{PSI-PR-13-03}
\title{The Higgs Singlet extension parameter space
in the light of the LHC discovery }
\vspace*{1.0truecm}
\author{Giovanni Marco Pruna}
\affiliation{Paul Scherrer Institute, CH-5232 Villigen PSI, Switzerland}
\affiliation{TU Dresden, Institut f\"ur Kern- und Teilchenphysik,
Zellescher Weg 19, D-01069 Dresden, Germany}
\author{Tania Robens}
\affiliation{TU Dresden, Institut f\"ur Kern- und Teilchenphysik,
Zellescher Weg 19, D-01069 Dresden, Germany}
\begin{abstract}
{\small \noindent
\ 
In this note we propose an overview on the current theoretical and
experimental limits on a Higgs singlet extension of the Standard
Model. We assume that the Boson which has
recently been observed by the LHC experiments is the lightest Higgs Boson of such model, while for the second Higgs Boson we consider a mass
range of $600\,\GeV\,\leq\,m_H\,\leq\,1\,\TeV$, where our model directly corresponds to a benchmark scenario of the heavy Higgs working group. In this light, we study the impact of
perturbative unitarity limits, renormalisation group equations
analysis and experimental constraints (electroweak precision tests,
measurements of the observed light Higgs coupling strength at the Large Hadron
Collider). We show that, in the case of no additional hidden sector contributions,
the largest constraints for higher Higgs masses stem from the assumption of perturbativity as
well as vacuum stability for scales of the order of the SM
metastability scale, and that the allowed mixing range is {\it severely restricted}. We discuss implications for current LHC searches in the singlet extension, especially the expected suppression factors for SM-like decays of the heavy Higgs. We present these results in terms of a global scaling factor $\kappa$ as well as the total width $\Gamma$ of the new scalar.
}
\end{abstract}
\maketitle
\tableofcontents

\newpage


\section{Introduction}
\label{Sec:Intro}
\noindent
The recent discovery of a new particle \cite{atlres,cmsres} which is in accordance with the scalar Boson from the Higgs mechanism \cite{Higgs:1964ia,Higgs:1964pj,Englert:1964et, Guralnik:1964eu, Kibble:1967sv} by the LHC experiments is one of the big breakthroughs in contemporary particle physics. If the discovered particle is indeed the Higgs Boson predicted from a SM-like Higgs-doublet sector, all its properties are completely determined by theory. Therefore, the current quest of the theoretical and experimental community is to establish whether the properties of such particle are in accordance with standard predictions, or  it is only a component of a more involved Higgs sector.  For this, all couplings as well as the spin structure of the new particle need to be severely tested.\\

In this work, we consider the simplest extension of the SM Higgs
sector, i.e. we add an additional singlet which is neutral under all
quantum numbers of the SM gauge groups
\cite{Schabinger:2005ei,Patt:2006fw} and acquires a vacuum expectation value (VEV) \cite{Barger:2007im, Bhattacharyya:2007pb, Dawson:2009yx, Bock:2010nz,Fox:2011qc, Englert:2011yb,
  Englert:2011us,Batell:2011pz, Englert:2011aa, Gupta:2011gd, Dolan:2012ac, Bertolini:2012gu,Batell:2012mj}. We assume that the heavy Higgs mass lies in the range $600\,\GeV\,\leq\,m_H\,\leq\,1\,\TeV$. While a second scalar state with a mass below 600 \GeV~ equally constitutes a viable scenario, we here focus on heavier additional resonances, in direct correspondence to one of the benchmark models of the heavy Higgs cross-section
working group \cite{hhwg,hhwg2,Heinemeyer:2013tqa}. This minimal setup can be interpreted as a limiting case for more generic BSM scenarios, as models with an additional gauge sectors (cf. e.g. \cite{Basso:2010jm}) or additional matter content (\cite{Strassler:2006im,Strassler:2006ri}). In our analysis, we combine the effects of several constraints: LHC bounds on the light Higgs signal strength, bounds from perturbative unitarity, electroweak (EW) parameters in terms of S,T, and U, and limits from perturbative running of the couplings. As a major result, we find that, for $m_H\,\gtrsim\,700\,\GeV$, especially the running of the couplings {\it severely restricts} the allowed parameter space of the model, leading to scaling factors in the percent range. In order to facilitate the comparison of our findings with results from the LHC experiments from searches in the heavy Higgs range, we express the bounds we obtain on the fundamental parameters of the theory in terms of a global suppression factor $\kappa$ for SM-like channels as well as the total width $\Gamma_H$ of the heavy Higgs, and exhibit regions which are allowed in the $\kappa,\,\Gamma$ plane. These can then directly related to LHC production cross sections at a $8$ and $14\,\TeV$ LHC.\\

This paper is organized as follows: In Section II, we briefly review the model setup. Section III is devoted to the investigation of the allowed parameter space taking all constraints into account. In Section IV, we comment on the impact of these limits on LHC observables. We summarize in Section V.



\section{The model}
\label{Sec:Model}
\noindent


\subsection{Potential and couplings}

In this paragraph we will shortly review our model: we enlarge the SM Higgs
sector with a further real Higgs singlet $\chi$, which is pure singlet under each gauge group of the SM \cite{Schabinger:2005ei,Patt:2006fw, Bowen:2007ia}.

The most general gauge-invariant and renormalisable scalar Lagrangian
is then:
\begin{equation}\label{lag:s}
\mathscr{L}_s = \left( D^{\mu} H \right) ^{\dagger} D_{\mu} H + 
D^{\mu} \chi D_{\mu} \chi - V(H,\chi ) \, ,
\end{equation}
with the scalar potential given by
\begin{eqnarray}\label{potential}\nonumber
V(H,\chi ) &=& -m^2 H^{\dagger} H - \mu ^2  \chi ^2 +
\left(
\begin{array}{cc}
H^{\dagger} H &  \chi ^2
\end{array}
\right)
\left(
\begin{array}{cc}
\lambda_1 & \frac{\lambda_3}{2} \\
\frac{\lambda_3}{2} & \lambda _2 \\
\end{array}
\right)
\left(
\begin{array}{c}
H^{\dagger} H \\  \chi^2 \\
\end{array}
\right) \\
\nonumber \\ 
&=& -m^2 H^{\dagger} H -\mu ^2 \chi ^2 + \lambda_1
(H^{\dagger} H)^2 + \lambda_2  \chi^4 + \lambda_3 H^{\dagger}
H \chi ^2,
\end{eqnarray}
where $x$ is the Vacuum Expectation Value VEV associated to the new Higgs field. {\cblack We here implicitely impose a $Z_2$ symmetry which forbids additional terms in the potential.} \\

To determine the condition for $V(H,\chi )$ to be bounded from below,
it is sufficient to study its behaviour for large field values,
controlled by the matrix in the first line of Eqn (\ref{potential}). Requiring such a matrix to be positive-definite gives the
conditions
\begin{eqnarray}
4 \lambda_1 \lambda_2 - \lambda_3^2 &>& 0 , \label{bound_pot}\\
\lambda_1, \lambda_2 &>& 0, \label{pos_pot}
\end{eqnarray}
{\cblack where the condition given by Eqn. (\ref{pos_pot}) corresponds to the requirement that the potential is bounded from below for large field values, while Eqn. (\ref{bound_pot}) guarantees that the extremum is indeed a local minimum.}\footnote{We give the exact derivation of the resulting eigenstates and the derivation in Appendix \ref{app:hpot} and here only cite the relevant results.}
Since the physical mass eigenvalues are gauge invariant, we define the
Higgs fields following the unitary-gauge prescription:
\begin{equation}\label{unit_higgs}
H \equiv
\left(
\begin{gathered}
0 \\
\frac{\tilde{h}+v}{\sqrt{2}}
\end{gathered} \right), 
\hspace{2cm}
\chi \equiv \frac{h'+x}{\sqrt{2}}.
\end{equation}

The explicit expressions for the scalar
mass eigenvalues are:
\begin{eqnarray}\label{mh1}
m^2_{h} &=& \lambda_1 v^2 + \lambda_2 x^2 - \sqrt{(\lambda_1 v^2 -
  \lambda_2 x^2)^2 + (\lambda_3 x v)^2}, \\
\label{mh2}
m^2_{H} &=& \lambda_1 v^2 + \lambda_2 x^2 + \sqrt{(\lambda_1 v^2 -
  \lambda_2 x^2)^2 + (\lambda_3 x v)^2},
\end{eqnarray}
where $h$ and $H$ are the scalar fields of definite masses
$m_{h}$ and $m_{H}$ respectively, with
$m^2_{h} < m^2_{H}$.

These eigenvalues are related to the following eigenvectors:
\begin{equation}\label{eigenstates}
\left(
\begin{array}{c}
h \\
H
\end{array}
\right) = \left(
\begin{array}{cc}
\cos{\alpha} & -\sin{\alpha} \\
\sin{\alpha} & \cos{\alpha}
\end{array}
\right) \left(
\begin{array}{c}
\tilde{h} \\
h'
\end{array}
\right),
\end{equation}
where $-\frac{\pi}{2} \leq \alpha \leq \frac{\pi}{2}$
fulfils\footnote{In all generality, the whole interval $0 \leq \alpha
  < 2\pi$ is halved because an orthogonal transformation is invariant
  under $\alpha \rightarrow \alpha + \pi$.}:
\begin{eqnarray}\label{sin2a}
\sin{2\alpha} &=& \frac{\lambda_3 x v}{\sqrt{(\lambda_1 v^2 -
    \lambda_2 x^2)^2 + (\lambda_3 x v)^2}}, \\
\label{cos2a}
\cos{2\alpha} &=& \frac{\lambda_2 x^2 - \lambda_1
  v^2}{\sqrt{(\lambda_1 v^2 - \lambda_2 x^2)^2 + (\lambda_3 x v)^2}}.
\end{eqnarray}
From Eqn. (\ref{eigenstates}), it is clear that the light (heavy) Higgs couplings to SM particles are now suppressed by $\cos\al\,(\sin\al)$.

From equations~(\ref{mh1})-(\ref{mh2})-(\ref{sin2a}), it is straightforward
to have:
\begin{eqnarray}\label{isomorphism}
\lambda_1&=&\frac{m_{h}^2}{2 v^2} + \frac{ \left(
m_{H}^2 - m_{h}^2 \right)}{2 v^2}\sin^2{\alpha} =
\frac{ m_{h}^2}{2v^2}\cos^2{\alpha} +  \frac{m_{H}^2}{2
v^2}\sin^2{\alpha}\nonumber \\
\lambda_2&=&\frac{m_{h}^2}{2 x^2} + \frac{ \left(
m_{H}^2 - m_{h}^2 \right)}{2 x^2}\cos^2{\alpha} =
\frac{ m_{h}^2}{2x^2}\sin^2{\alpha} +  \frac{m_{H}^2}{2
x^2}\cos^2{\alpha} \nonumber \\
\lambda_3&=&\frac{ \left(
m_{H}^2 - m_{h}^2 \right)}{ 2vx}\sin{(2\alpha)}.
\end{eqnarray}

In summary, the heavy Higgs is a ``twin'' version of the light Higgs with rescaled couplings to the matter contents of the SM. In fact, the only novel channel with respect to the light Higgs case is $H\to hh$.
The decay width $\Gamma$ and coupling strength $\mu'$ of the $H\,\rightarrow\,h\,h$ decay are  \cite{Schabinger:2005ei, Bowen:2007ia}:
\begin{eqnarray}\label{eq:muprime}
&&\Gamma\,\lb H\,\rightarrow\,h\,h \rb\,=\,\frac{|\mu'|^2}{8\,\pi\,m_{H}}\,\sqrt{1-\,\frac{4\,m^2_{h}}{m_{H}^2}},\nonumber\\
&&\mu'\,=\,-\frac{\lambda_3}{2}\,\lb x\,\cos^3\,\al+v\,\sin^3\,\al \rb+\,\lb \lambda_3-3\,\lambda_1 \rb\,v\,\cos^2\al\,\sin\,\al\,+\,\lb \lambda_3-3\,\lambda_2 \rb\,x\,\cos\,\al\,\sin^2\,\al.\nonumber\\
&&
\end{eqnarray}

We here briefly discuss the behaviour of $|\mu'|$ when $x$ and $\sin\al$ are varied: from Eqn. (\ref{eq:muprime}), it is clear that\footnote{If $x\,\gg\,v$, we can approximate $\mu'\,\approx\,\lam_3\,x\,\cos\al\lb \sin^2\al-\frac{1}{2}\cos^2\al \rb-3\,\lam_1\,v\,\cos^2\al\sin\al\,+\,\mO\lb\frac{v}{x}\rb$. We then have $\mu'(\sin\al)\,=\,-\mu'(-\sin\al)$. If the terms $\sim\,\mO(x^{-1})$ cannot be neglected, they introduce a positive/ negative contribution to $|\mu'|$ depending on the sign of $\sin\al$.}
 $|\mu'|^2(|\sin\al|)\,>\,|\mu'|^2(-\sin\al)$. The difference is more pronounced as $x$ is increased. In addition, for a fixed value of $\sin\al>0 (\sin\al\,<0)$, $|\mu'|$ decreases (increases) constantly for increasing $x$. These features will become important in the discussion of the experimental and theoretical constraints in the next sections.\\

The model investigated here implies a global suppression factor for all SM-like couplings for the light/ heavy resonance respectively, determined by the additional parameters of the Higgs sector. We briefly want to comment on this feature. For example, if the apparent enhancement in the $h\,\rightarrow\,\gamma\,\gamma$ decay channel of the light Higgs had persisted, it might have rendered further studies of the model futile, at least on the level of a leading order analysis. However, recent results for the measurement of this branching ratio are in good agreement (within $\lesssim\,1.5\,\sigma$) with SM predictions \cite{ATLAS-CONF-2013-012,CMS-PAS-HIG-13-001}. Therefore, as long as a relative overall light Higgs coupling strength $\mu\,\lesssim\,1$ is not experimentally excluded, our model constitutes a viable extension of the SM Higgs sector. 

\subsection{Number of free parameters}
\noindent
Our simple singlet extension model has in principle 5 free parameters on the Lagrangian level
\begin{\eqn*}
\lambda_1,\,\lambda_2,\,\lambda_3,\,v,\,x.
\end{\eqn*}
The coupling parameters $\lambda_i$ are related to the masses and the effective mixing according to Eqns. (\ref{mh1}),(\ref{mh2}), and (\ref{sin2a}), and we obtain the independent parameters
\begin{\eqn}\label{eq:pars}
m_h,\,m_H,\,\alpha,\,v,\,x.
\end{\eqn}
Moreover, we will reexpress $x$ by $\tan\beta$ according to
\begin{\eqn*}
\tan\beta\,=\,\frac{v}{x}
\end{\eqn*}
to accomodate for standard notation in models with exended Higgs sectors.
If we assume the vacuum expectation value of the Higgs doublet value
to be Standard Model-like such that $v\,\sim\,246\,\GeV$, and equally
set the Higgs mass of the light Higgs to
$m_h\,=\,125\,\GeV$, we are left with
three independent parameters $m_H,\,\alpha,\,\lb x/\tan\be \rb$. All
results in the following sections will be given in dependence on these
variables. In this work, we restrict the range of the Singlet VEV to
$x\,\in\,[100\,\GeV;\,10\,\TeV]$, leading to $\tan\be\,\in\,[0.025;2.46]$.





\section{Theoretical and experimental bounds on the Higgs singlet extension}
\noindent
In this section, we will discuss the current theoretical and
experimental limits on the singlet extension model. We here consider{\color{green}:}
\begin{itemize}
\item{}limits from perturbative unitarity,
\item{}limits from EW precision data in form of the $S,\,T,\,U$ parameters,
\item{}perturbativity constraints on the couplings, as well as conditions on a potential which is bounded from below,
\item{} limits from measurements of the light Higgs signal strength, 
\item{}limits from perturbativity of the couplings as well as vacuum stability up to a certain scale $\mu_\text{run}$, where we chose $\mu_\text{run}\,\sim\,10^{10}\,\GeV,\;\,10^{19}\,\GeV$ as benchmark points.
\end{itemize}

In this chapter, we will investigate the parameter space
$(\sin\al,\,\tan\be)$, while keeping $m_H$ fixed; however, in order to
demonstrate the effects of the partial-wave treatment of perturbative
unitarity, we will equally comment on the highest possible mass of the
heavy Higgs $m_{H,\text{max}}$ in this parameter space, including exclusion bounds from electroweak precision data using $m_{H,\text{max}}$. We discuss all limits separately in the following subsections.

\subsection{Limits from perturbative unitarity}

\label{Sec:bounds}

Tree-level perturbative unitarity  \cite{Lee:1977eg,Luscher:1988gc} puts a constraint the Higgs masses of our theory via a relation on the partial wave amplitudes $a_J(s)$ of all possible $2\,\rightarrow\,2$ scattering processes:

\begin{eqnarray}\label{condition}
|\textrm{Re}(a_J(s))|\leq \frac{1}{2}.
\end{eqnarray}

In the high energy limit, $\sqrt
s\rightarrow \infty$, only the $a_0$ partial wave amplitude does not
vanish, instead it approaches a value depending only on $m_{h}$,
$m_{H}$, $\alpha$ and $x$. Therefore, by applying the condition in
eq.~(\ref{condition}), we can obtain several different (correlated)
constraints on the Higgs masses and mixing angle, i.e., we can find the
$m_{h}$-$m_{H}$-$\alpha$ subspace in which the perturbative
unitarity of the theory is valid up to any energy scale. We therefore studied the unitarity constraints in our model by
calculating tree-level amplitudes for all two-to-two processes\footnote{Calculations where actually carried out with the vectors Bosons being replaced by the corresponding Goldstone Bosons following the equivalence theorem \cite{Chanowitz:1985hj}.} 
 $X_1\,X_2\,\rightarrow\,Y_1\,Y_2$, with $(X_1,X_2),\,(Y_1,Y_2)\,\in\,(W^+\,W^-,ZZ,hh,hH,HH)$  in terms of the
mixing angle between the two physical Higgs fields and their masses. 
Then, we calculated the normalized $5$-dimensional bosonic scattering matrix and we imposed the condition of Eqn.~(\ref{condition}) to each of its eigenvalues (the largest in modulus gives the best constraint).
Note that, in accordance with \cite{Bowen:2007ia}, the constraint based on generic unitarity considerations (cf. e.g. \cite{Englert:2011yb}) for the heavy Higgs of $m_H\,\lesssim\,700\,\GeV$ is much loosened\footnote{This result is also confirmed in \cite{Basso:2010jt}, where a similar scenario is investigated.}.\\

\begin{figure}[!tb]
\begin{minipage}{0.49\textwidth}
 \includegraphics[width=1.1\textwidth ]{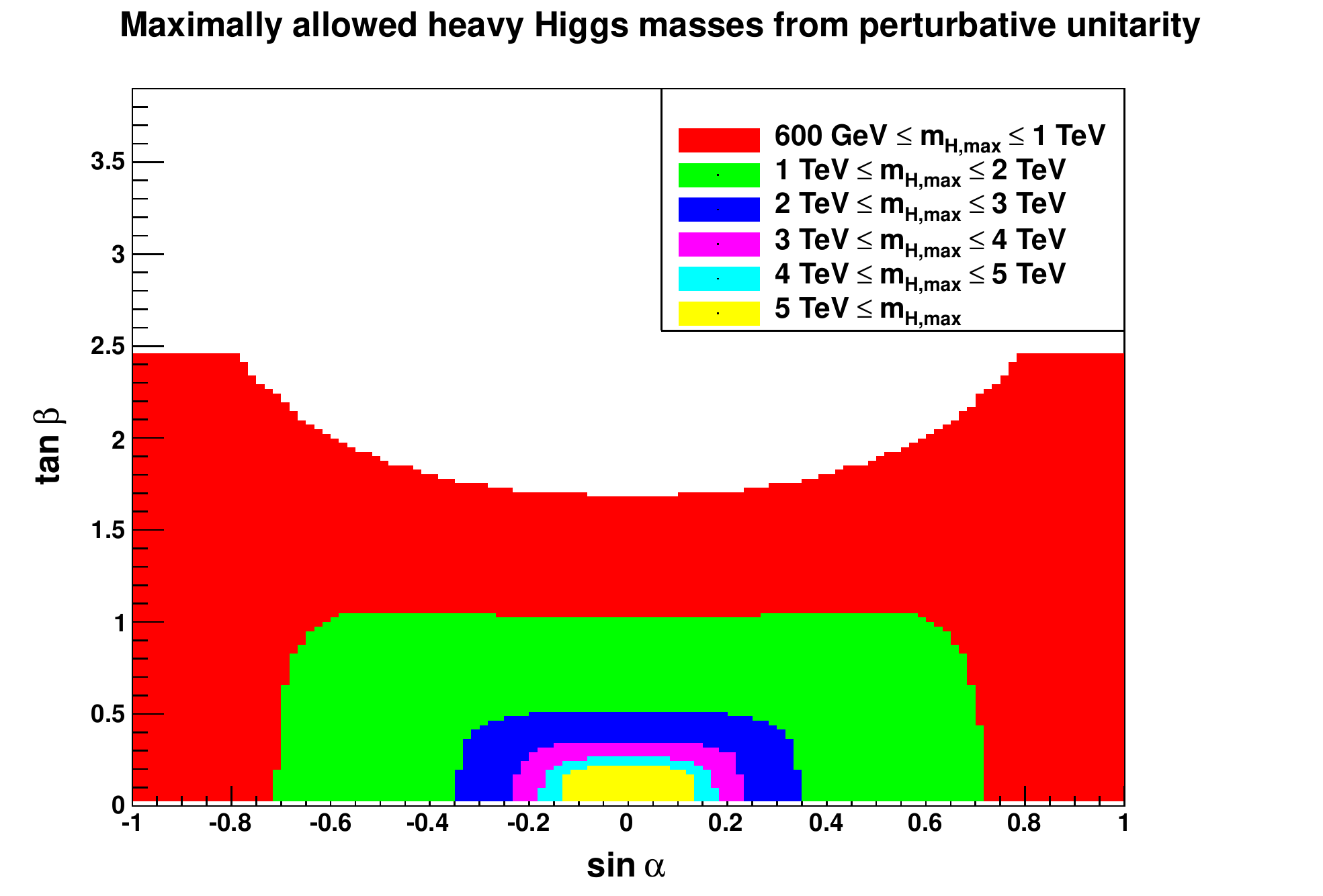}
\end{minipage}
\begin{minipage}{0.49\textwidth}
\includegraphics[width=1.1\textwidth ]{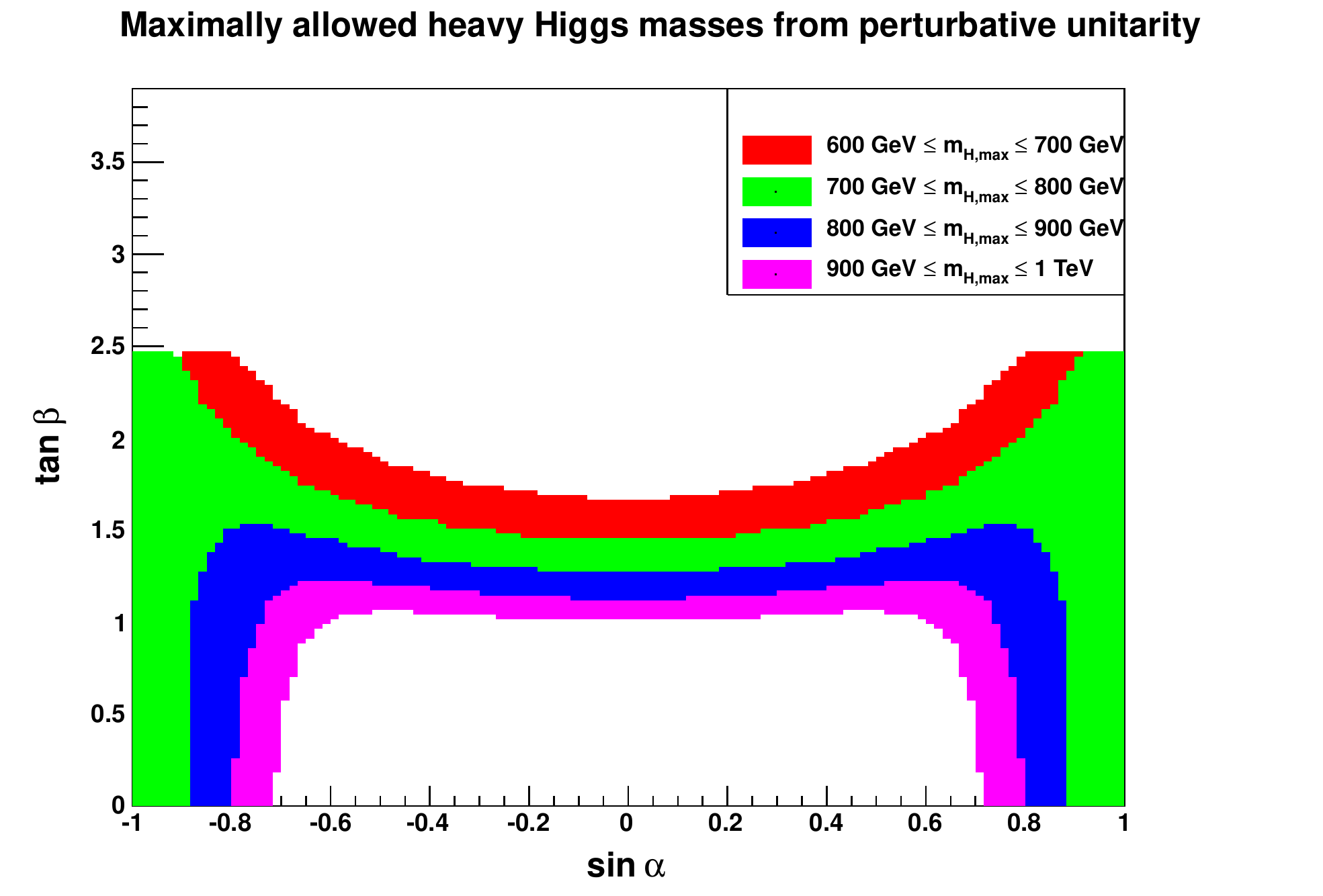}
\end{minipage}
\caption{ \label{fig:uplim} Allowed mixing ranges for different maximal heavy Higgs masses $m_{H,\text{max}}$
in dependence of the mixing angle $\al$ as well as ratio of the VEVs
$\tan\beta$. {\sl LEFT:} Heavy Higgs masses above
$(1,\,2,\,3,\,4)\,\TeV$ are excluded in the (red, green, blue,
magenta, yellow) area (from $\sin\al\,=\,-1$ to $\sin\al\,=\,1)$. The
region exluded for small $\sin\al$  and $\tan\beta\,\gtrsim\,2$ leads
to $m_{H,\text{max}}<600\,\GeV$. The
most important constraints on the upper limit in the small mixing
region stem from scattering processes involving heavy Higgs pairs (not
shown here). {\sl RIGHT:} Zoom into the region where $m_{H,\text{max}}\,\leq\,1\,\TeV$.}
\end{figure}

Figure \ref{fig:uplim} shows the regions in
parameter space which are still allowed after limits from perturbative
unitarity only. We found that for small mixing angles within our scan range, the most dominant contribution stems from scattering processes involving only heavy Higgses. For $\sin\al\,\sim\,0$, the scattering matrix becomes approximately block diagonal with a SM block and the decoupled $HH\to HH$ element, and the latter gives the unitarity limits on the singlet VEV, i.e.\footnote{This boundary is in fact stronger than perturbativity of the coupling alone, which leads to $\tan^2\be\,\leq\,\frac{8\,\pi\,v^2}{m_H^2}$ for $\sin\al\,=\,0$.}
\begin{\eqn*}
\tan^2\be\,\leq\,\frac{16\,\pi\,v^2}{3\,m_H^2}\,+\,\mO\lb \al \rb\;\;\;\text{for } a_0(HH\,\rightarrow\,HH)\,\leq\,0.5;
\end{\eqn*}
if $\tan\be$ is decreased accordingly, this boundary can therefore be fulfilled any heavy Higgs mass. For small, but non-zero mixing angles $|\sin\al|\,\sim\,0.02$ and $\tan\be\,\lesssim\,0.1$ , upper limits for the allowed maximal heavy Higgs mass can reach up to $35\,\TeV$. We found that generically, heavy Higgs scattering processes dominate for $\tan\be\,\gtrsim\,1.5$; if $\tan\be$ is decreased, and for non-zero mixing, gauge Boson scattering becomes equally important. However, in most cases the whole $5\,\times\,5$ scattering matrix involving all partial wave contributions  needs to be considered, and an approximation considering a single dominant process cannot give a valid predicition of the upper limit on the allowed heavy Higgs mass.


\subsection{Limits from electroweak precision data}

Constraints from EW precision data are incorporated using the $S,\,T,\,U$ parameters \cite{Peskin:1991sw, Hagiwara:1994pw}, which parametrize deviations from the SM predictions and thereby render constraints on new physics from higher order corrections stemming from BSM contributions.

We here follow \cite{Espinosa:2012im}, which cites the
 values for the EW parameters as
\begin{\eqn*}
S\,=\,0.00\,\pm\,0.10,\;T\,=\,0.02\,\pm\,0.11,\;U\,=\,0.03\,\pm\,0.09
\end{\eqn*}
and equally used $m_h\,=\,125\,\GeV$ as an input value for the
calculation of the SM reference values where
$S_\text{ref}=T_\text{ref}=U_\text{ref}\,\equiv\,0$. As a cross check,
we have compared results from our code with the values for $S,\,T,\,U$
for all benchmark points specified
in \cite{Bowen:2007ia}, and found agreement with small variations on
the $10\,\%$ level (we used\footnote{In \cite{Bowen:2007ia}, the actual value of
  the top mass which was used is not given. Variations for
  $m_t\,\in\,[170.5;173.5]$ did not significantly change our results.}
$(m_\text{top},m_h)\,=\,(173.5,150)\,\GeV$).   To accomodate for this slight disagreement, we decreased the allowed regions for $S,\,T,\,U$ to
\begin{\eqn*}
S\,=\,0.00\,\pm\,0.095,\;T\,=\,0.02\,\pm\,0.105,\;U\,=\,0.03\,\pm\,0.085
\end{\eqn*}
in our scans, and use as input variables \cite{Beringer:1900zz}
\begin{\eqn*}
\hat{s}_Z\,=\,0.2313,\,\al_s(M_Z)\,=\,0.120,\,m_t\,=\,173\,\GeV.
\end{\eqn*}
We then use \cite{Gupta:2012mi}
\begin{\eqn*}
X_\text{tot}\,=\,\cos^2\al\,X(m_h)\,+\,\sin^2\al\,X(m_H)
\end{\eqn*}
with $X\,\in\,\left[ S,T,U \right]$. Note that this approach neglects suppression of the couplings in all but the leading order, and equally does not take the $H\,h\,h$ couplings into account which can appear in higher order corrections including heavy Higgses running in the loops\footnote{See also \cite{Dawson:2009yx} for a generic calculation with multiple scalar extensions of the SM.}. However, as we will argue below, EW precision data basically poses no constraint on the parameter space after all other restrictions have been taken into account. Of course, a more detailed analysis would be desireable here, and is in the line of future work.
In our approximation, the constraints
basically rule out values of $|\sin\al|\,\geq\,0.5-0.7$ depending on
$\tan\be$, where the strongest constraints here come from the
$T$-parameter. $U$ does not pose any additional constraints. \\

This closes our discussion of scans using maximally allowed heavy Higgs masses from perturbative unitarity. We found that within our scan range the maximally allowed Higgs masses are $\mO(35\,\TeV)$ for small mixing angles and that, using these maximal Higgs masses, EW precision data give additional constraints in the large mixing regions. For $m_{H,\text{max}}\,\leq\,1\,\TeV\,(2\,\TeV),\,|\sin\al|\,\lesssim\,0.6\,(0.5)$. For $m_{H,\text{max}}\,\geq\,2\,\TeV$, EW precision data give no additional constraints to our model.\\

In the following, we will fix the Higgs mass to $m_H\,\in\,[600,\,\GeV; 1\,\TeV]$, and equally include the measurement of the light Higgs signal strength $|\mu|$ as well as vacuum stability and perturbativity of the couplings up to a metastable scale/ the Planck scale. We will see that indeed these latter requirements are much more stringent than  EW precision data and the light Higgs measurements and render severe constraints on the parameter space of our model.

\subsection{Constraints from the signal strength of the light Higgs} 

If we want to accomodate for the light Higgs measurements \cite{cmsres,atlres}, we need to take into account the limits on the maximally allowed value of $|\sin\al|$ from the overall signal strength $|\mu|$. In general, we have\footnote{In fact, loop-induced couplings like the $h\,\rightarrow\,\gamma\gamma$ branching ratio in principle call for a more refined treatment, cf. e.g. discussion in \cite{Bertolini:2012gu}. However, the corresponding corrections are generally on the sub-permill level, and can therefore safely be ignored in our simple limit-setting. For fitting procedures, on the other hand, such a more complex coupling structure needs to be taken into account.}
\begin{\eqn*}
\mu\,\equiv\,\frac{\sigma_\text{BSM}(m_h)}{\sigma_\text{SM}(m_h)}\,=\,\frac{\cos^4\,\al\,\Gamma_\text{SM}(m_h)}{\cos^2\al\,\Gamma_\text{SM}(m_h)\,+\,\sin^2\,\al\,\Gamma_\text{hid}(m_h)},
\end{\eqn*}
where $\Gamma_\text{SM/ hid} (m_h)$ denote the decay widths of the light Higgs in the SM/ a possible hidden sector. We here constrain ourselves\footnote{Taking decays in the hidden sector into account additionally reduces the allowed mixing range. To understand this, consider the case that $\Gamma_\text{SM}\,=\,\Gamma_\text{hid}$; in this case, the constraint is strengthened to $\cos^4\al\,\geq\,0.95$, leading to $\sin\al\,\leq\,0.17$. The case in which $\Gamma_\text{hid}\,=\,0$ is therefore the best case scenario.} to cases where $\Gamma_\text{hid}\,=\,0$; then, the above equation leads to $\mu\,=\,\cos^2\al$. 

The values measured by the LHC experiments  \cite{cmsres,atlres} then render \footnote{The official ATLAS fit for a $126\,\GeV$ Higgs are given by \cite{atlres} $\mu\,=\,1.4\,\pm\,0.3$, and from CMS for a $125\,\GeV$ Higgs as \cite{cmsres}  $\mu\,=\,0.87\,\pm\,0.23$. Being conservative, we consider $\mu\,\geq\,0.95$, where we take the fact
  into account that the model considered here cannot accomodate for $\mu\,>\,1$. We also assumed that the errors of the ATLAS and CMS measurements are completely uncorrelated. {\cblack Newer values \cite{ATLAS-CONF-2013-034,CMS-PAS-HIG-13-005}  do not significantly change this result}. See also \cite{Bertolini:2012gu} for a best fit result for this model. }
\begin{eqnarray*}
|\sin\,\alpha|\,\in\,[0;0.23]&\text{from}&\mu\,\in\,[0.95;1].
\end{eqnarray*}
These limits on the
  measurement of the $125/ 126 \,\GeV$ Higgs Bosons coupling strength are in
  fact much more stringent than  EW precision observables.

\begin{figure}[!tb]
\begin{minipage}{0.49\textwidth}
\includegraphics[width=1.1\textwidth ]{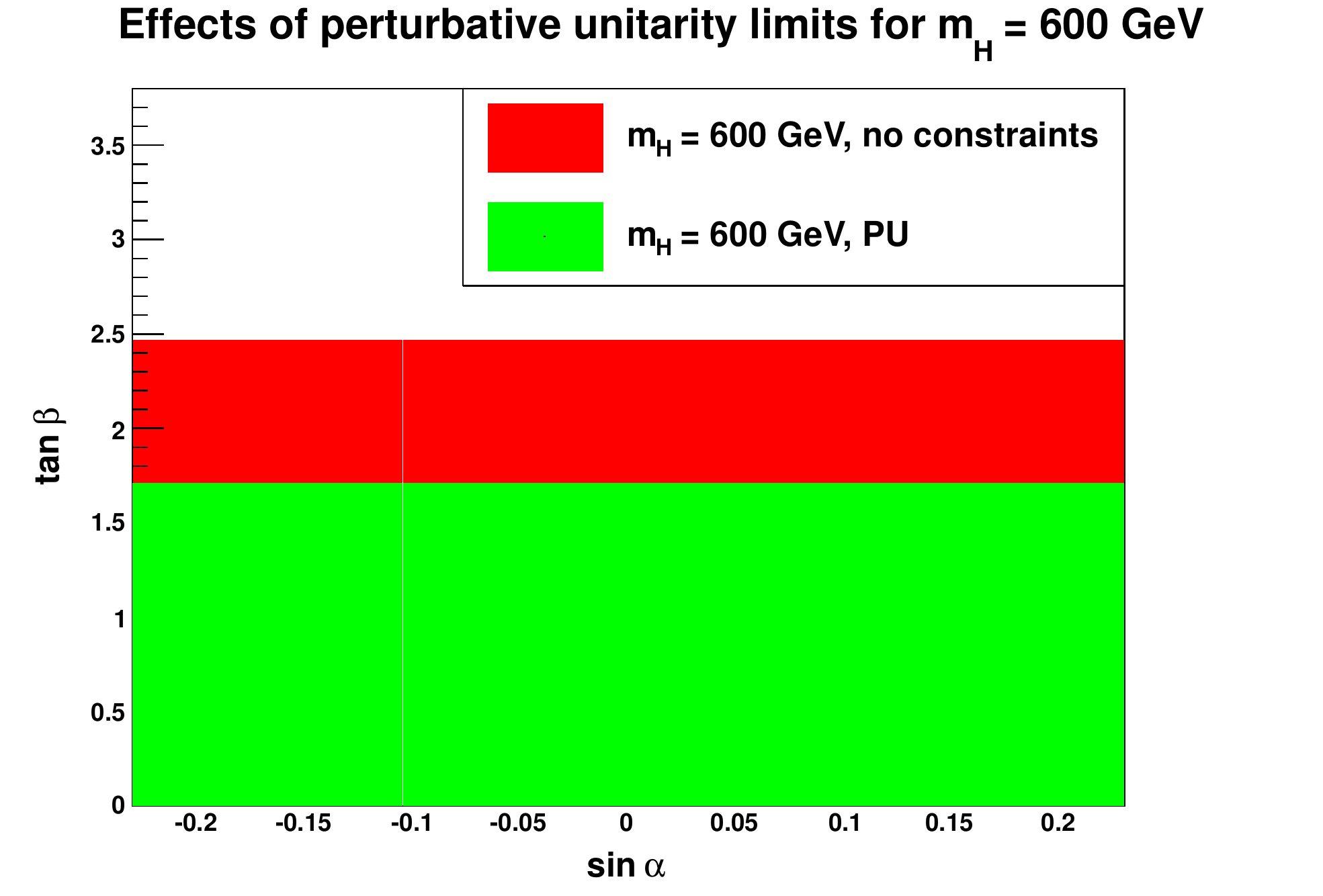}
\end{minipage}
\begin{minipage}{0.49\textwidth}
 \includegraphics[width=1.1\textwidth ]{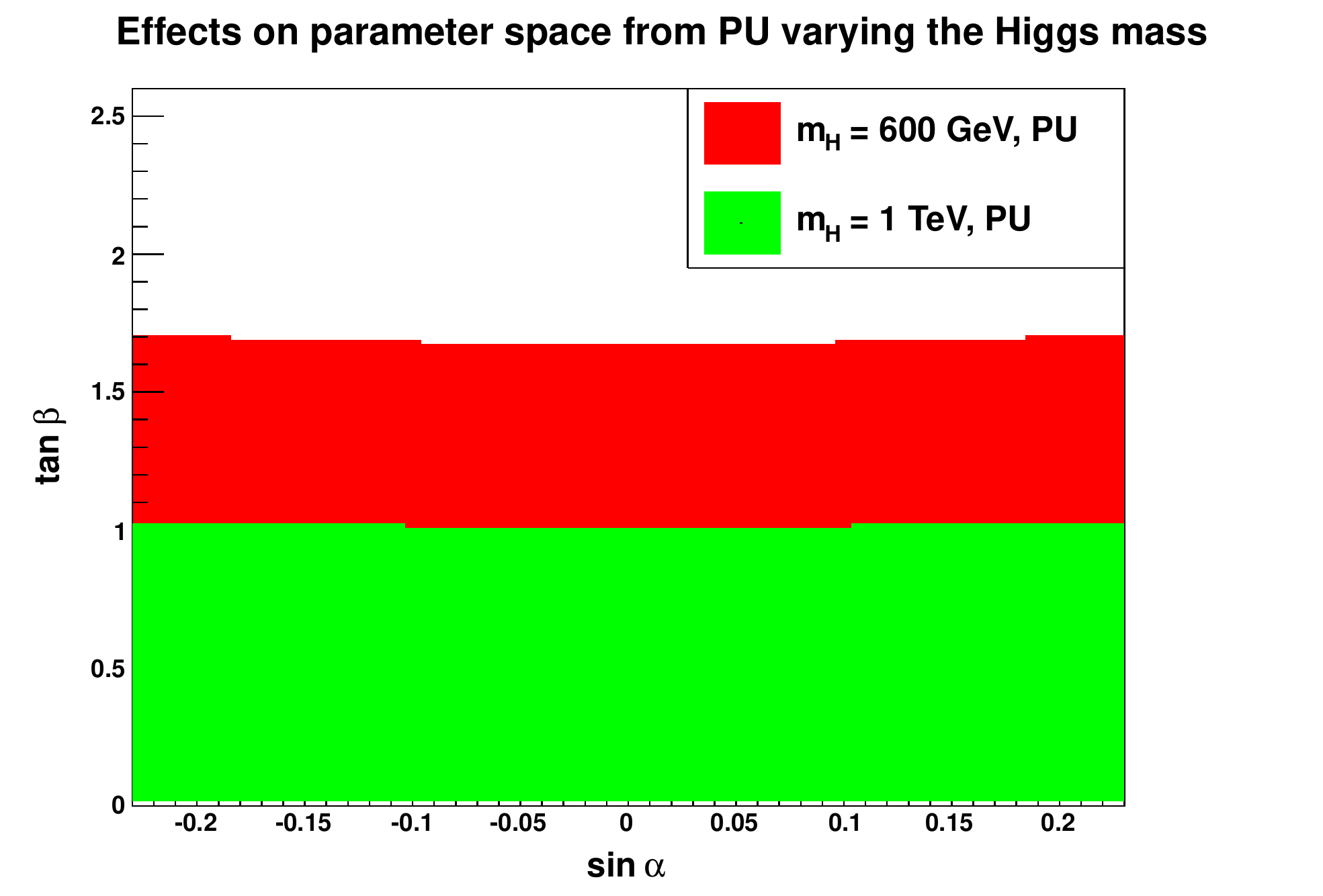} 
\end{minipage}
\caption{ \label{fig:600}{\sl LEFT:} Allowed regions in parameter space where $m_H\,=\,600\,\GeV$. The only restriction comes from PU, which gives an upper limit on $\tan\be$. {\sl RIGHT:}  Allowed parameter space for a $600\,\GeV$ as well as $1\,\TeV$ heavy Higgs. As before, the only limit in parameter space comes from perturbative unitarity, which fixes the upper allowed bound of $\tan\,\beta$ to 1.9 and 1.0 respectively. }

\end{figure}

 Figure \ref{fig:600} shows the limits for  $m_H\,=\,600\,\GeV$ and $m_H\,=\,1\,\TeV$ respectively. The only constraint arises here from perturbative unitarity, which sets an upper limit on $\tan\be$ in both cases. This is generically due to a large $\lam_2$ value in these regions of parameter space. In accordance with the behaviour of $\lam_2\,\sim\,m_H^2$ for fixed $(\sin\al,\,\tan\be)$ values, we equally observe that the coupling gets larger for larger $m_H$ values, leading to a decrease in the upper limit of $\tan\beta$. Most of the parameter space ruled out by perturbative unitarity would however be equally excluded by the requirement of perturbativity of $\lam_2$ at the  EW scale, as discussed below.

\subsection{Limits from perturbativity and vacuum stability}
\label{sec:rgerun}
We equally consider vacuum stability as well as perturbativity of the Higgs potential couplings up to a certain scale $\mu_\text{run}$. Vacuum stability follows from Eqn. (\ref{pos_pot}), while perturbativity of the couplings leads to the requirement that
\begin{\eqn*}
\lambda_{1,2}(\mu_\text{run})\,\leq\,4\,\pi,\;|\lambda_3(\mu_\text{run})|\,\leq\,4\,\pi.
\end{\eqn*}
At the electroweak scale, we found that these conditions pose no additional constraints on the allowed parameter space of the model, when limits from the light Higgs signal strength and perturbative unitarity are taken into account. If we neglect perturbative unitarity limits, the upper allowed values of $\tan\be$ following from perturbativity of the couplings alone are $2.05\,(1.24)$ for $m_H\,=\,600\,\GeV\,(1\,\TeV)$ (for $\sin\al\,=\,0$) respectively, which slightly enhances the allowed $\tan\be$ ranges. Before considering the running of the couplings, we can therefore say that
\begin{itemize}
\item perturbative unitarity alone indeed allows for heavy Higgses in the 30 \TeV~ range
\item the strongest constraints considered so far, when the experimental results for the light Higgs signal strength are taken into account, stem from perturbative unitarity.
\end{itemize}

We now discuss limits from perturbativity up to $\mu_\text{run}$, where we use the running parameter
$t\,=\,\ln\lb\frac{\mu_\text{run}^2}{v^2} \rb$ such that $t\,=\,0$ for
$\mu_\text{run}\,=\,v$. {\cblack We here impose the constraint given by Eqn. (\ref{bound_pot}) at all energies. Note that in a strict sense this is not required for vacuum stability; for positive $\lam_3$ values, fulfilling Eqn. (\ref{pos_pot}) is sufficient, cf. e.g. the discussions in \cite{Lebedev:2012zw,Belanger:2012zr}. However, as we require perturbative unitarity up to arbitrary high scales, we also demand that the process of electroweak symmetry breaking remains the same and that therefore the minimum of the potential is indeed positioned at the VEVs of the two fields; this approach has e.g. been followed in \cite{Basso:2010jm}. We will briefly comment on the effects of releasing such a condition on the collider observables in Section IV.}\\

In the following discussion, we mostly focus on
$m_H\,=\,600\,\GeV,\,\mu_\text{run}\,=\,2.7\,\times\,10^{10}\,\GeV\,(t\,=\,37)$, but will equally give results for
$m_H\,=\,1\,\TeV$ and $\mu_\text{run}\,=\,10^{19}\,\GeV$. In the end of the
discussion, we will comment on the generic changes for a higher Higgs
mass or the requirement of perturbativity and vacuum stability at
higher scales. For the sake of the argument, we will temporarily neglect the measurement of the light Higgs signal strength and consider mixing angles $|\sin\al|\,\leq\,0.49$ in the discussion of RGE running effects at the low scale, in order to exemplify the generic effects on the parameter space. The signal strength measurement will however be included again in the discussion of collider observables in Section IV.  \\

The renormalization group equations for this model are given by \cite{Bowen:2007ia,Basso:2010jm}
\begin{eqnarray*}
\frac{d}{dt}\lambda_1&=&\frac{1}{16\,\pi^2}\left\{\frac{1}{2}\lam_3^2+12\,\lam_1^2+6\,\lam_1\,y_t^2-3\,y_t^4-\frac{3}{2}\lam_1\,\lb 3\,g^2+g_1^2 \rb\,+\,\frac{3}{16}\left[ 2\,g^4+\lb g^2+g_1^2\rb^2  \right]  \right\},\\
\frac{d}{dt}\lam_2&=&\frac{1}{16\,\pi^2}\left[ \lam_3^2+10\,\lam_2^2 \right],\\
\frac{d}{dt}\lam_3&=&\frac{1}{16\,\pi^2}\,\lam_3\,\left[ 6\,\lam_1+4\,\lam_2+2\,\lam_3+3\,y_t^2-\frac{3}{4}\lb 3\,g^2+g_1^2 \rb \right],
\end{eqnarray*}
where $y_t$ is the (equally running) top Yukawa-coupling and $g,\,g_1$ are the running couplings of the SM gauge groups. For the decoupling case as well as for cross check for the running of the gauge couplings, for which we chose the analytic solution at one loop, we reproduced the results in \cite{Degrassi:2012ry}, where the SM breakdown scale following the one-loop treatment here was at $t\,=\,36$ corresponding to a scale $\mu_\text{run}\,\sim\,1.6\,\times\,10^{10}\,\GeV$. By choosing a benchmark value of $\mu_\text{run}\,=\,2.7\,\times\,10^{10}\,\GeV\,(t\,=\,37)$, we are able to investigate which regions of parameter space are still allowed at a scale which slightly exceeds the SM breakdown scale; in this sense, our model can solve (or at least postpone) the metastability problem of the SM. Even with such stringent constraints, substantially large regions of parameter space are still allowed. In addition, the requirement of vacuum minimization at such scales complies with the requirement of perturbative unitarity for $\sqrt{s}\,\rightarrow\,\infty$.\\

\begin{figure}
\begin{minipage}{0.49\textwidth}
\includegraphics[width=\textwidth]{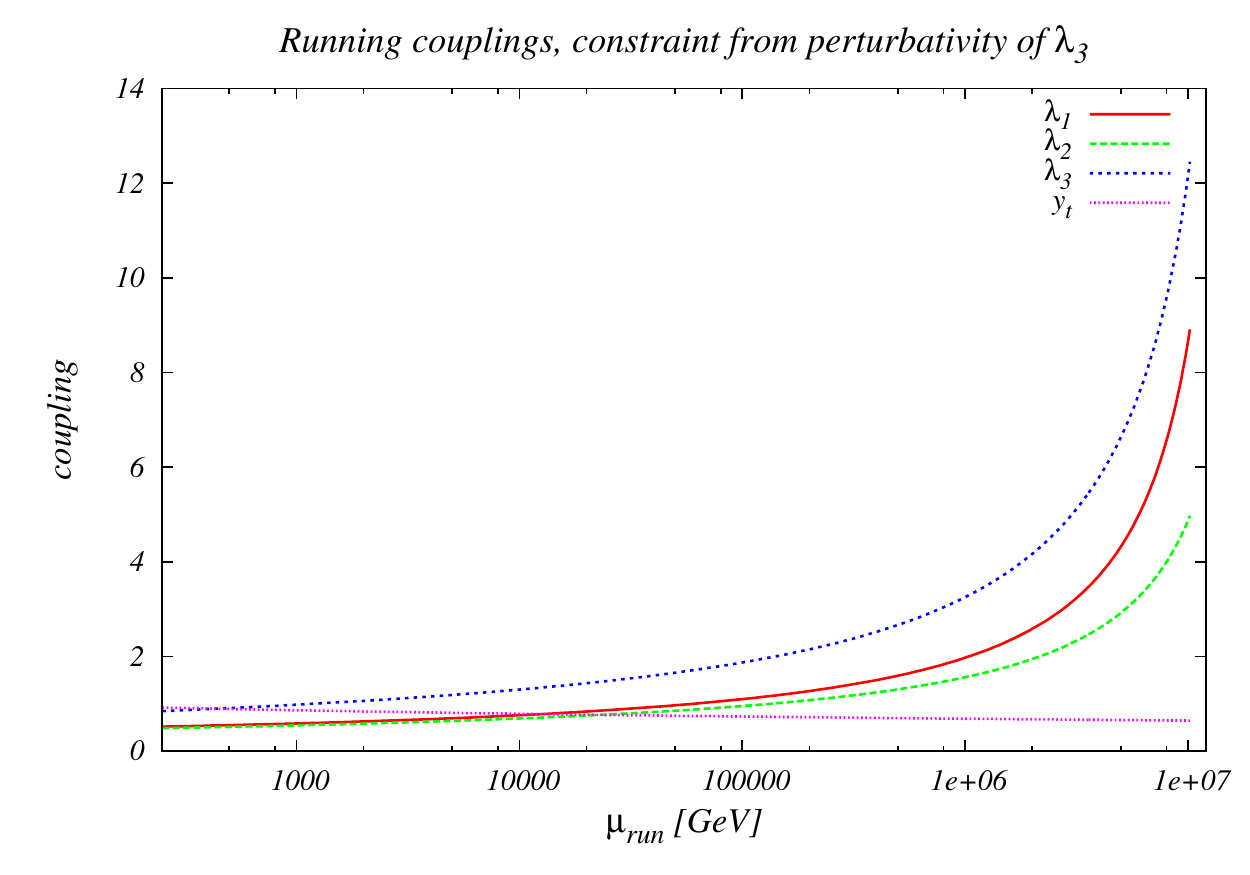}
\end{minipage}
\begin{minipage}{0.49\textwidth}
\includegraphics[width=\textwidth]{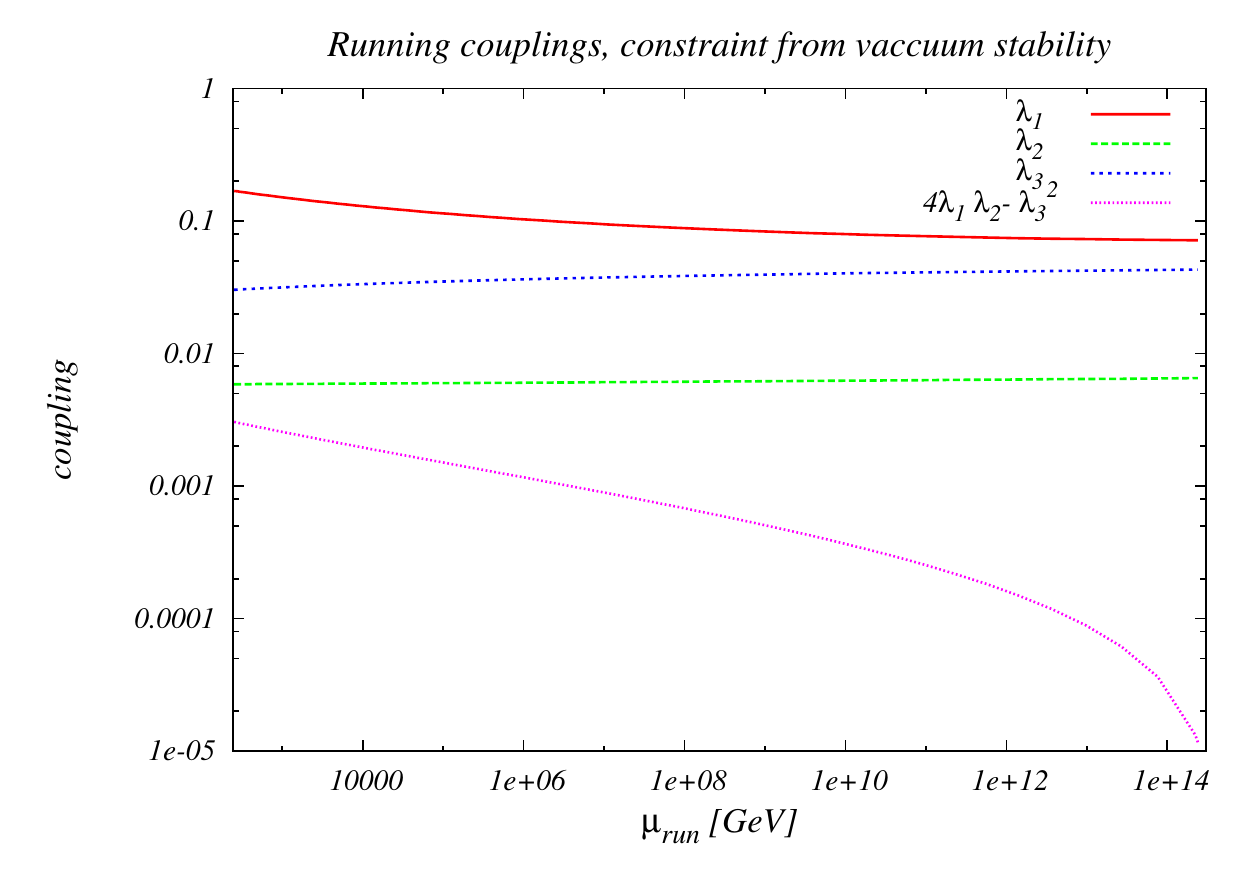}
\end{minipage}
\caption{\label{fig:rge_run} Example for RGE running: {\sl LEFT:} $\lam_3$ becomes non-perturbative at $ \mu_\text{run}\sim \mO(10^7\,\GeV)$, with $\sin\al\,=\,0.37,\,\tan\be\,=\,0.43$, {\sl RIGHT:} Example for region where $4\,\lam_1\,\lam_2\,-\lam_3^2$ only varies marginally  over large scale ranges, with $\sin\al\,=\,0.12,\,\tan\be\,=\,0.04$.}
\end{figure}

We found that the strongest constraints from a phenomenological
viewpoint, i.e. upper limits on the allowed mixing angle, actually stem
from perturbativity of Higgs self-couplings $\lam_1,\,\lam_2$; for $\mu_\text{run}\,=\,2.7\,\times\,10^{10}\,\GeV\,(t\,=\,37)$ and low $m_H$, we found the requirement that
$|\sin\al|\lesssim\,0.3$. This poses a much stronger constraint than electroweak precision
tests. In the following, we discuss limits from perturbativity as well as vacuum stability in more detail:





\begin{itemize}
\item {\sl Perturbativity of $\lam_1$ and upper limit on $|\sin\al|$}\\
The strongest constraint on large mixing angles stems from the running
of $\lam_1$. For $\tan\be\,\lesssim\,0.1$ and large mixing angles
\begin{\eqn*}
\lam_2,\,|\lam_3|\,\ll\,\lam_1\,\sim\,y_t
\end{\eqn*}
 at the electroweak scale,
so the $\beta$ function of $\lam_1$ is positive. In this case,
$\lam_1$ quickly grows and approaches the upper limit of
$4\,\pi$ (eg for $\sin\al\,\sim\,0.49,\,\tan\be\,\sim\,0.025$, this is
reached for the relatively low scale of $\mu_\text{run}\,\sim\,350\,\TeV$). This remains the dominant effect until $\tan\be\,\gtrsim\,0.36$, where $\lam_2$ starts to rise more quickly. From the running of $\lam_1$, we obtain $|\sin\al|\,\lesssim\, 0.3\,(0.2)$ for running up to $\mu_\text{run}\,=\,2.7\,\times\,10^{10}\,\GeV$ (the Planck scale) for a 600 \GeV~ Higgs; for 1 \TeV, these values change to $0.2\,(0.12)$.
\item {\sl Perturbativity of $\lam_2$ and upper limit on $\tan\be$}\\
For $\tan\be\,\gtrsim\,0.36$, the most dominant constraint comes from the running of $\lam_2$ in almost all regions of parameter space\footnote{In the region where $|\sin\al|\,\gtrsim\,0.26$, $\lam_3$ running sets in as well, cf. discussion below.}. Generically, a good estimate of the limits can be obtained by considering the zero-mixing case and $\tan\be\,\gtrsim\,1$: we then have
\begin{\eqn*}
\lam_2\,\gg\,\lam_1,\,\lam_3.
\end{\eqn*}
In this case, it is easy to estimate the maximal value of $\tan\be$ allowed such that $\lam_2\,=\,4\,\pi$. The corresponding $\beta$-function can be reduced to 
\begin{\eqn*}
\frac{d\lambda_2}{dt}\,\sim\,\frac{5}{8\,\pi^2}\lam_2^2,
\end{\eqn*} 
which has the solution
\begin{\eqn*}
\lam_2(t)\,=\,\frac{\lam_2(t=0)}{1-\frac{5}{8\,\pi^2}\,t\,\lam_2(t=0)}.
\end{\eqn*}
Requiring $\lam_2(t)\,\leq\,\lam_\text{max}$
then leads to
\begin{\eqn*}
t\,\leq\,\frac{\lam_\text{max}-\lambda(t=0)}{\frac{5}{8\,\pi^2}\,\lam_2(t=0)\,\lam_\text{max}}.
\end{\eqn*}
As $\lam_2(t=0)\,\sim\,\frac{1}{x^2}$, this translates to a lower limit on $x$
\begin{\eqn*}
x_\text{min}^2(t)\,=\,\frac{m_H^2}{2}\,\left[\frac{1}{\lam_\text{max}}+\frac{5}{8\,\pi^2}\,t \right]
\end{\eqn*}
(here we set $\sin\al\,=\,0$) and therefore $\tan\be_\text{max}\,=\,\frac{v}{x_\text{min}}$.
Inserting explicit values for $m_H\,=\,(600,\,\GeV;1\,\TeV)$ for $t\,=\,37\,(76)$ renders $\lb\tan\be\rb_\text{max}\,=\,\lb 0.37\,(0.26);\;0.22\,(0.15)\rb$.
These values agree with our numerical findings.
\item {\sl Perturbativity of $\lam_3$ and restriction in the large $\tan\be$/ large $\sin\al$ region}\\
In a small region for $\tan\be\,\sim\,0.4$ and large positive mixings, 
\begin{\eqn*}
\lam_3\,\gtrsim\,\lam_1,\,\lam_2,
\end{\eqn*}
which corresponds to the transition between $\lam_1$ and $\lam_2$
dominance. In this region, all couplings evolve similarly fast up to
high scales. As an example, we show the running of all Higgs sector as well as the top Yukawa
coupling for a point in this part of parameter space in Figure \ref{fig:rge_run}.
\item {\sl First vacuum stability condition ($\lam_1\,\geq\,0$)  and minimal mixing angle $|\sin\al|$}\\
For small (or 0) mixings, this is the well-known metastability problem of the SM Higgs\footnote{See \cite{Gunion:1989we} for a generic introduction, and \cite{Degrassi:2012ry} for recent work.} with a low mass of $125\,\GeV$. In our scan, the couplings becomes negative at a scale $t\,=\,36$, which corresponds to roughly $\mu_\text{run}\,\sim\,1.6\,\times\,10^{10}\,\GeV$. For small mixing angles $|\sin\al|\,\lesssim\,0.001$, the problem persists. There is no significant change from this limit for raising the Higgs mass to $1\,\TeV$ \footnote{\cblack We want to mention that larger heavy Higgs masses allow for $\lam_1\,\geq\,0$ for running up to arbitrary scales, cf. e.g. \cite{Lebedev:2012zw,EliasMiro:2012ay}. However, the mass of the second Higgs Boson is typically much above the LHC reach in the according setup. We thank O. Lebedev for useful discussions regarding this point.}. 
\item {\sl Third vaccum stability condition ($4\,\lam_1\,\lam_2\,\geq\,\lam_3^2$) and minimal mixing angle $|\sin\al|$}\\
For $0.001\,\leq\,|\sin\al|\,\leq\,0.4$ and $\tan\be\,\lesssim\, 0.4$, the third vacuum stability condition
\begin{\eqn*}
4\,\lam_1\,\lam_2\,-\,\lam_3^2\,\geq\,0
\end{\eqn*}
poses the largest constraints.
For a $600\,\GeV$ Higgs mass, mixing angles
between $0.001$ and $0.04$ are excluded, where for larger $\tan\be$
the upper limits are slightly less stringent. For a Higgs mass of 1
\TeV, this region is decreased to $0.02$. Increasing $\sin\al$, the
transition into the allowed region comes from an enhanced value of
$\lam_1$ at the low scale; in
this case, the limiting value is again the perturbativity of
$\lam_1$. Note that in parts of the parameter space  $\lam_2,\lam_3$
only change marginally; in this case, there is a very fine interplay
between the rise of the absolute values of $\lam_2,\,\lam_3$ and the
rapid decrease of $\lam_1$, so including additional orders in the
running might change these bounds, leading to a larger allowed
region. An example for such a ``slow-running'' point is given in
Figure \ref{fig:rge_run}. If we want to prevent this fine-tuning over
large scales, we could e{\color{green}.}g{\color{green}.} allow for slightly negative values of
$4\,\lam_1\,\lam_2-\lam_3^2$; opening up the condition such that
$4\,\lam_1\,\lam_2\,-\lam_3^2\,\geq\,-0.001$ leads to
$(\sin\al)_\text{min}\,\sim\,0.015\,(0.01)$ for a 600 \GeV~ (1 \TeV) Higgs mass.
In priniciple, this area of parameter space would need a more detailed investigation. However, this region is phenomenologically difficult test, and the most important limits are indeed the ones from perturbativity on the maximal allowed mixing, so we will not investigate this in more detail in this work.
\end{itemize}

\subsubsection*{Summary of RGE effects}

\begin{figure}[!tb]
\begin{minipage}{0.49\textwidth}
\includegraphics[width=1.1\textwidth]{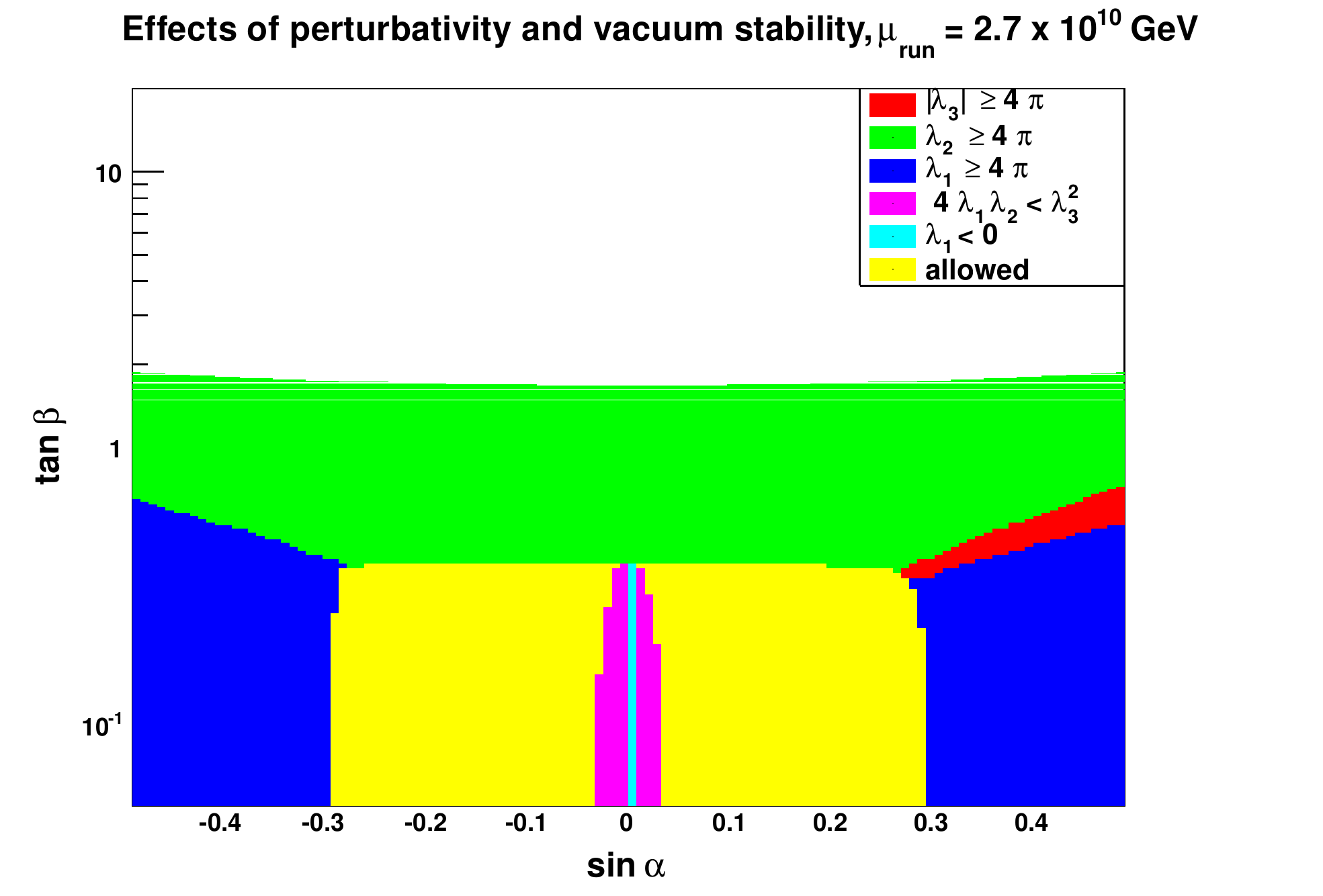}
\end{minipage}
\begin{minipage}{0.49\textwidth}
\includegraphics[width=1.1\textwidth]{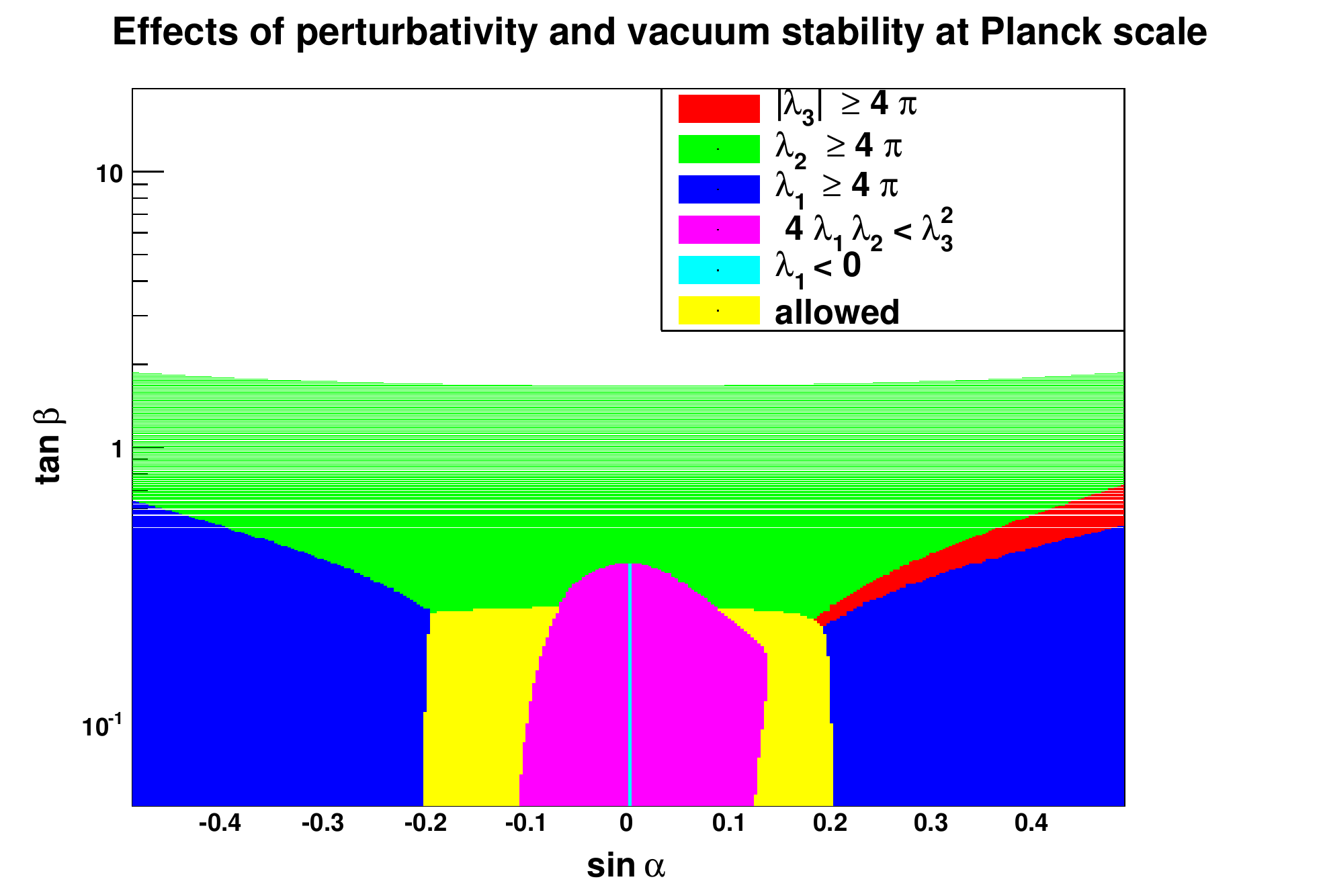}
\end{minipage}
\caption{\label{fig:mh600_scales} Limits for $m_H\,=\,600\,\GeV$ at $\mu_\text{run}\,=\,2.7\,\times\,10^{10}\,\GeV\,(t\,=\,37)$ (left) as well as Planck scale (right).  We here consider $|\sin\al|\,\leq\,0.49$; the experimental limit is given by $|\sin\al|\,\lesssim\,0.23$}
\end{figure}

\begin{figure}
\begin{minipage}{0.49\textwidth}
\includegraphics[width=1.1\textwidth]{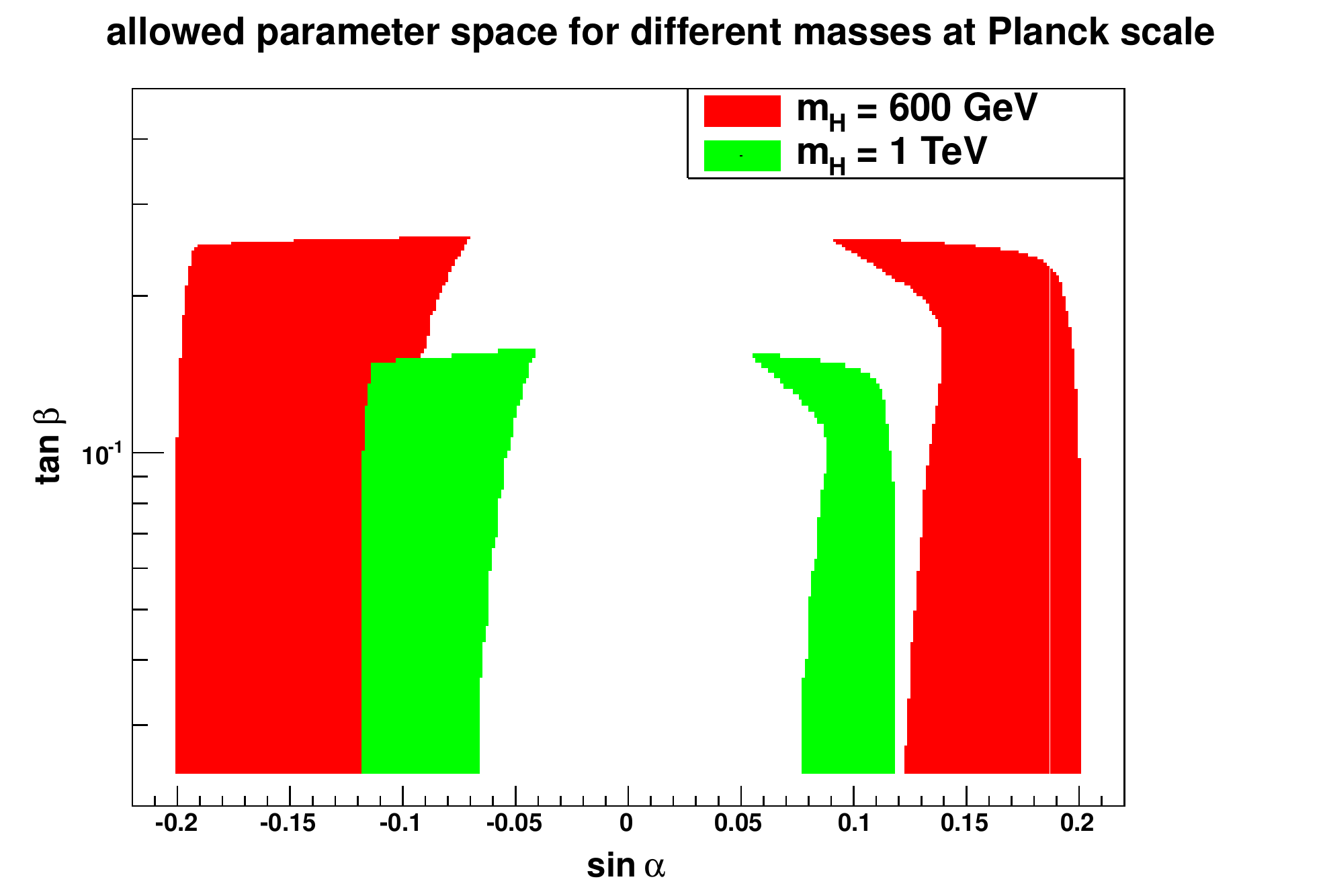}
\end{minipage}
\begin{minipage}{0.49\textwidth}
\includegraphics[width=1.1\textwidth]{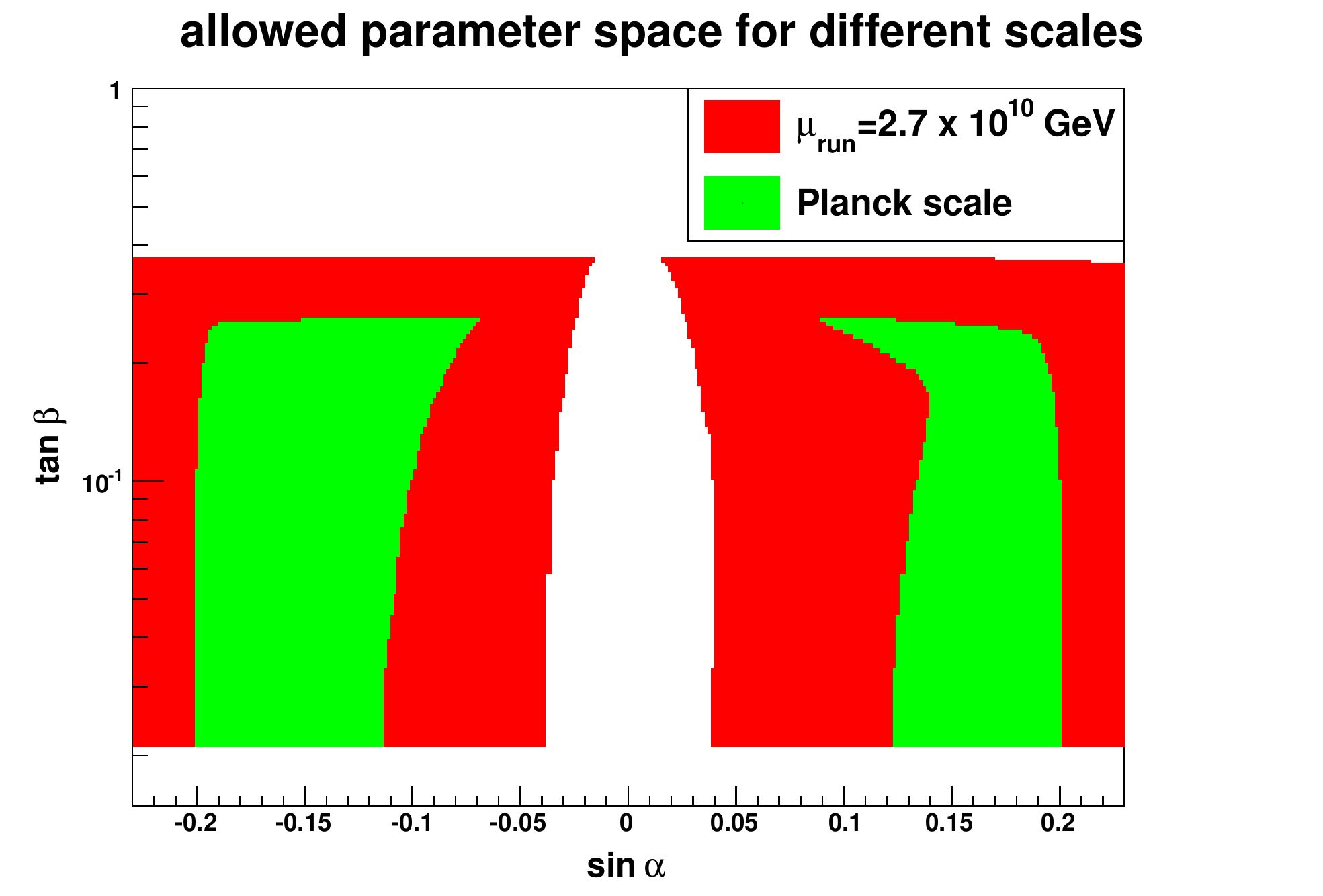}
\end{minipage}
\caption{\label{fig:massscalecomp} Limits at Planck scale for different $m_H$ (left) as well as limits for $m_H\,=\,600\,\GeV$ at different scales (right). SM signal strength measurements have been neglected.}
\end{figure}

In this subsection, we will first summarize the results for a 600 \GeV~ Higgs at a running scale corresponding to $\mu_\text{run}\,=\,2.7\,\times\,10^{10}\,\GeV\,(t\,=\,37)$ and then discuss variations of the heavy Higgs mass and consequences when going to a higher scale. In Figure
\ref{fig:mh600_scales}, we show the allowed parameter space for 600 \GeV~ Higgs mass both at the
low ($\mu_\text{run}\,=\,2.7\,\times\,10^{10}\,\GeV,\,t\,=\,37$) and the Planck scale. As discussed above, the largest constraints on
large mixing angles are given by running of $\lam_1$ and $\lam_2$ for
low/ high $\tan\be$ regions respectively, while generally $\tan\be\,\gtrsim\,0.37\,(0.26)$ is excluded by $\lam_2$ running at the low (high) scale. Additionally, small $|\sin\al|$ values are generically excluded from requiring vaccuum stability. The minimal/ maximal values for $|\sin\al|$ are $\sim\,0.035/ 0.3 (0.1/ 0.2)$ at $\mu_\text{run}\,=\,2.7\,\times\,10^{10}\,\GeV$ (the Planck scale).
Although the above discussion focuses on a Higgs mass of $600\,\GeV$, the characteristics of the respective limits remain the same if the mass or the scale of the running are increased. We observe the following effects:
\begin{itemize}
\item{} {\sl Raising the heavy Higgs mass while keeping the scale fixed}
  leads to a reduction of the maximal allowed mixing angle, which
  stems from the perturbativity of $\lam_1$, as well as a decrease of the
  allowed maximal value of $\tan\be$ from perturbativity of
  $\lam_2$. However, on the other hand smaller mixings are still
  allowed. This is due to a larger $\lam_2$ value at the
   EW scale, which prevents a fast decrease of $4\,\lam_1\,\lam_2$: this equally holds for larger $\lam_3$. Even for negative $\beta_{\lam_1}$ function values at low scales, the growth of $\lam_2,\,\lam_3$ can prevent $\lam_1$ from becoming negative\footnote{E.g., such a point is given by $\sin\al\,=\,0.1,\,\tan\be\,=\,0.05$, which is excluded (allowed) for $m_H\,=\,600\,\GeV\, (1\,\TeV)$ by requiring stability up to the Planck scale.}. In general, the allowed region shrinks and equally moves to smaller mixing angles and $\tan\be$ values. The effects are displayed in Figure \ref{fig:massscalecomp}, where we compare the allowed parameter space at the Planck scale for a $600\,\GeV$ as well as $1\,\TeV$ heavy Higgs mass.
\item{} {\sl Raising the scale while keeping the  Higgs mass fixed} has similar effects: the maximal allowed mixing angle area is further restricted; generally, the allowed region is shrinking and moving to smaller minimally allowed $\tan\be$ values, cf. Figures \ref{fig:mh600_scales} and \ref{fig:massscalecomp}.
\end{itemize}

\begin{figure}[!tb]
\begin{minipage}{0.49\textwidth}
\includegraphics[width=1.1\textwidth ]{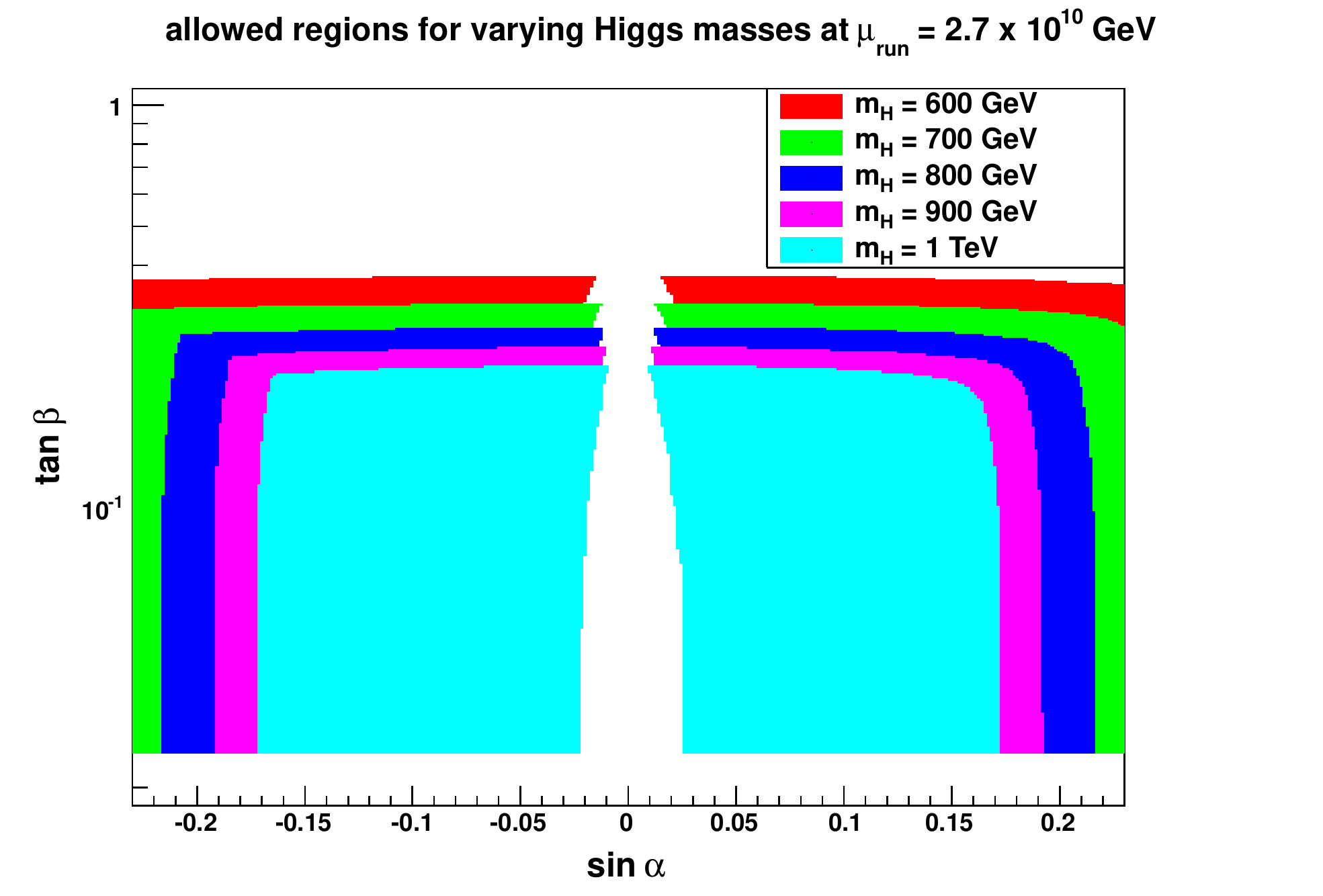}
\end{minipage}
\begin{minipage}{0.49\textwidth}
\includegraphics[width=1.1\textwidth]{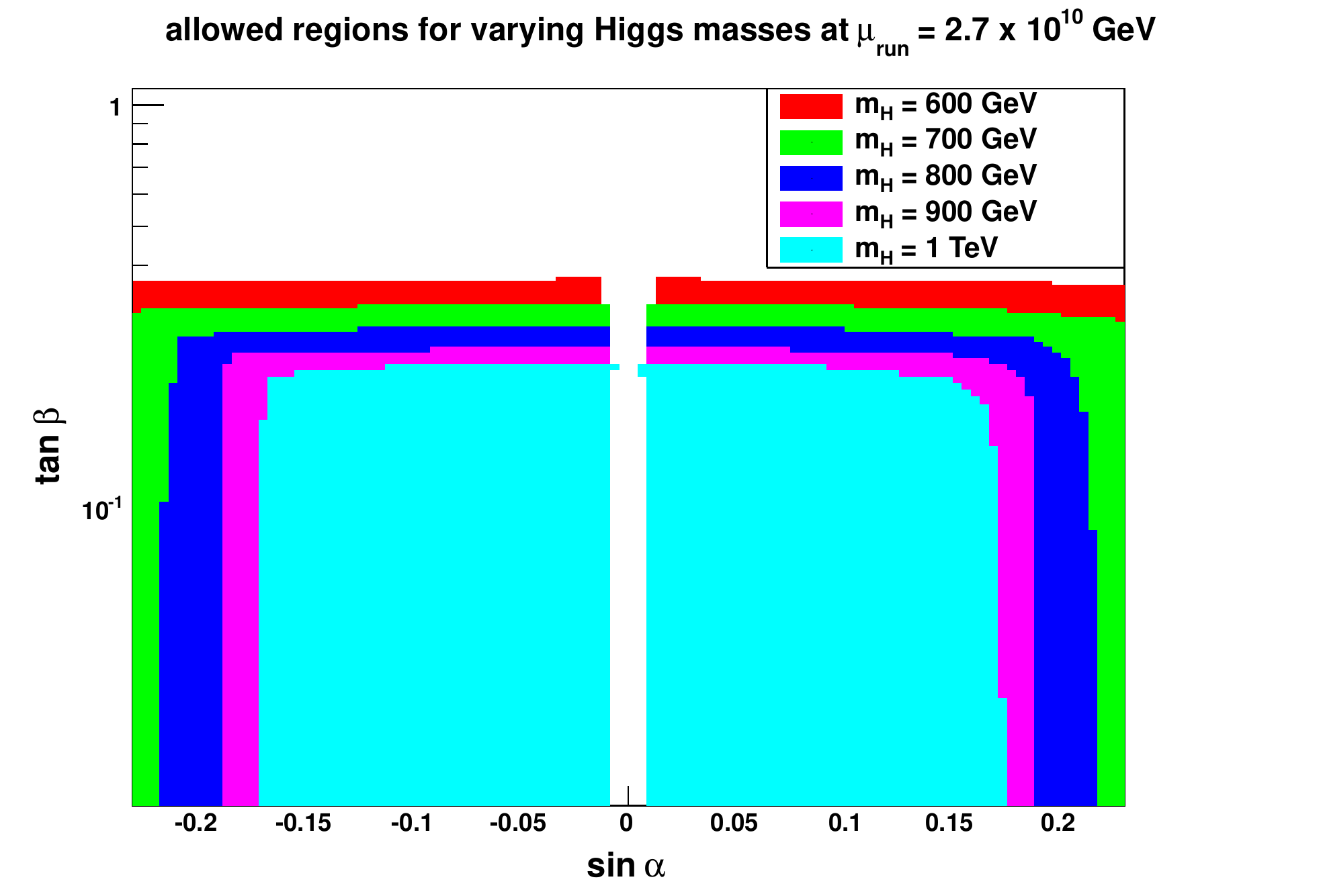}
\end{minipage}
\caption{\label{fig:alllow} Contour plots for different (fixed) $m_H$ at $\mu_\text{run}\,=\,2.7\,\times\,10^{10}\,\GeV$. In the right plot, the third vacuum stability condition $4\,\lam_1\,\lam_2\,\geq\,\lam^2_3$ has been neglected, leading to a slightly enhanced region for small $|\sin\al|$ values. SM signal strength measurement has been taken into account. }
\end{figure}

\begin{figure}[!tb]
\begin{minipage}{0.49\textwidth}
 \includegraphics[width=1.1\textwidth]{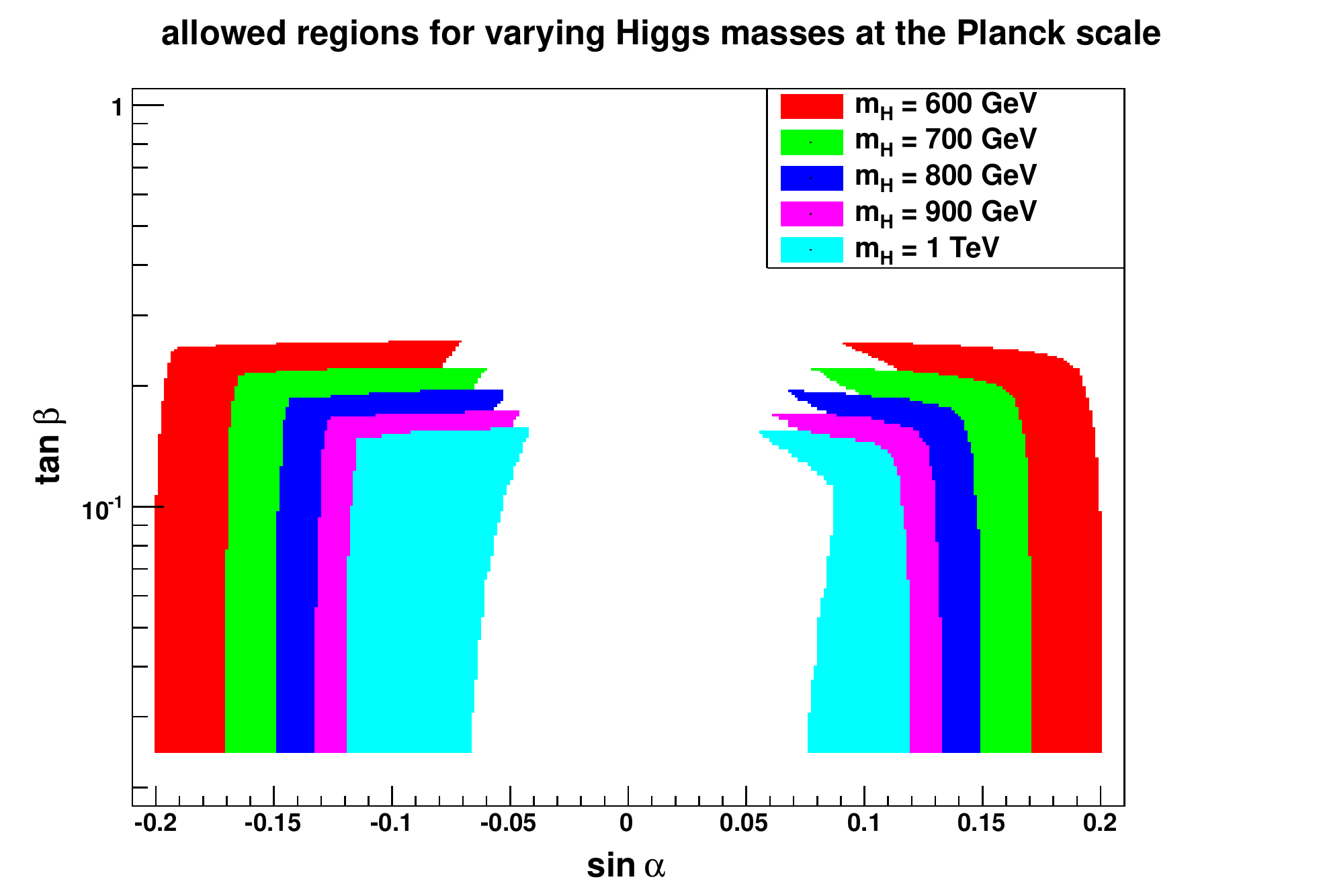}
\end{minipage}
\begin{minipage}{0.49\textwidth}
 \includegraphics[width=1.1\textwidth]{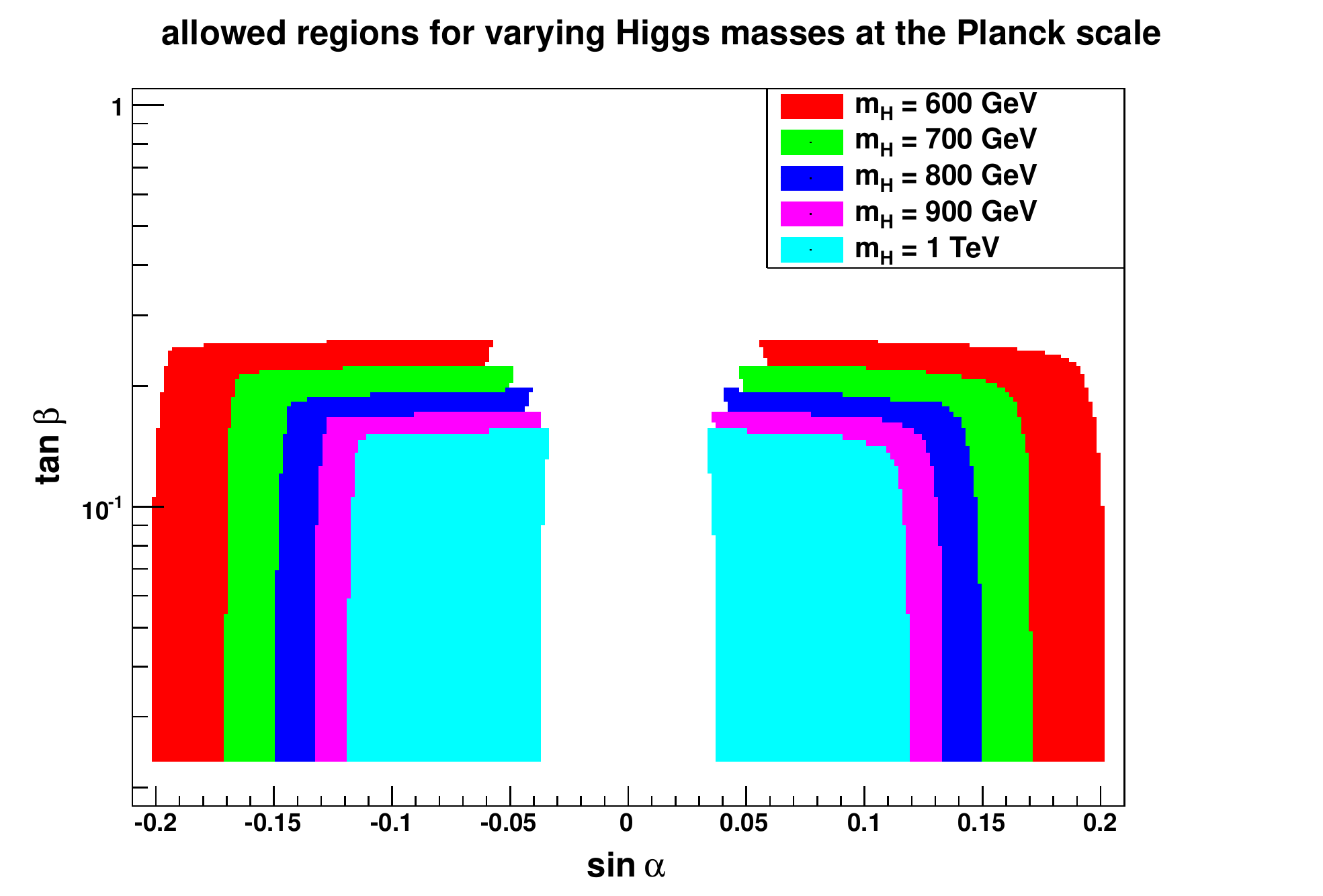}
\end{minipage}
\caption{\label{fig:allplanck} As Fig. \ref{fig:alllow}, but for validity up to the Planck scale.}
\end{figure}

In Figures \ref{fig:alllow} and \ref{fig:allplanck}, we present the results of our scans including all limits in terms of  contour plots for the allowed areas at $\mu_\text{run}\,=\,2.7\,\times\,10^{10}\,\GeV$ as well as the Planck scale for $m_H\,=\,600,\,700,\,800,\,900,\,1000\,\GeV$, with numerical values summarized in Tables \ref{tab:smbd3} and \ref{tab:planck}. As discussed above, the validity of the third vacuum stability condition, i.e. 
\begin{\eqn*}
4\,\lam_1\,\lam_2\,-\lam_3^2\,\geq\,0
\end{\eqn*}
using NLO precision only might be questioned, so we equally present results where this is neglected. In general, this opens up the parameter space for even smaller mixing angles. We then take the constraints from vacuum stability following $\lam_1$ running as a conservative lower limit. 

\begin{table}
\begin{\eqn*}
\begin{array}{c||c|c|c} 
m_H&|\sin\al|_\text{min,37}&|\sin\al|_\text{max,37}&(\tan\be)_\text{max, 37}\\ \hline\hline
600&0.035&0.23&0.37\\
700&0.030&0.23&0.31\\
800&0.024&0.21&0.28\\
900&0.019&0.19&0.25\\
1000&0.016&0.17&0.22\\
\end{array}
\end{\eqn*}
\caption{\label{tab:smbd3} Allowed parameter ranges for varying Higgs masses at the low ($\mu_\text{run}\,=\,2.7\,\times\,10^{10}\,\GeV$) scale; $|\sin\al|_\text{min,max}$ taken at $\tan\be=0.15$. $x\,\leq\,1\,\TeV$ fixes the lowest $\tan\be$ value to $0.025$. For $m_H\,\lesssim\,700\,\GeV$, the maximal allowed mixing angle results from the measurement of the light Higgs signal strength.}
\end{table}
\begin{table}
\begin{\eqn*}
\begin{array}{c||c|c|c} 
m_H&|\sin\al|_\text{min,Planck}&|\sin\al|_\text{max,Planck}&(\tan\be)_\text{max, Planck}\\ \hline\hline
600&0.104&0.2&0.26\\
700&0.086&0.17&0.22\\
800&0.074&0.15&0.20\\
900&0.064&0.13&0.17\\
1000&0.055&0.12&0.15\\
\end{array}
\end{\eqn*}
\caption{\label{tab:planck} Allowed parameter ranges for varying Higgs masses at the Planck scale;$|\sin\al|_\text{min,max}$ taken at $\tan\be=0.08$ and at $\sin\al\,<\,0$. As before, the minimal value of $\tan\be$ is determined by the scan range.}
\end{table}

\section{Translation to collider observables}
\noindent
The parameter space presented in the last subsection translates into two different observables at colliders:
\begin{itemize}
\item the generic suppression of the {\sl production} of the heavy Higgs; this is given by $\sin^2\al$,
\item suppression of the {\sl SM decay modes} of the heavy Higgs. Here, we have to take into account that the additional mode
\begin{\eqn*}
H\,\rightarrow\,h\,h
\end{\eqn*}
leading to a further reduction of the SM-like branching ratios.
\end{itemize}
The total width of the heavy Higgs is then modified to
\begin{\eqn*}
\Gamma_\text{tot}\,=\,\sin^2\al\,\Gamma_\text{SM, tot}+\Gamma_{H\,\rightarrow\,h\,h}.
\end{\eqn*}
Following the observables tested by the experiments, we therefore consider
\begin{\eqn*}
\kappa,\;\Gamma_\text{tot},
\end{\eqn*}
with the global SM-scaling factor defined as
\begin{\eqn*}
\kappa\,\equiv\,\frac{\sigma_\text{BSM}\,\times\,\text{BR}_\text{BSM}}{\sigma_\text{SM}\,\times\,\text{BR}_\text{SM}}\,=\,\frac{\sin^4\al\,\Gamma_\text{tot,SM}}{\Gamma_\text{tot}}
\end{\eqn*}
In analogy with the above definition, we also introduce a scaling factor $\kappa'$ which parametrizes the $H\,\rightarrow\,h\,h$ decay:
\begin{\eqn*}
\kappa'\,\equiv\,\frac{\sigma_\text{BSM}\,\times\,\text{BR}_\text{BSM}}{\sigma_\text{SM}}\,\,=\,\frac{\sin^2\al\,\Gamma_{H\,\rightarrow\,h\,h}}{\Gamma_\text{tot}},
\end{\eqn*}
where $\kappa+\kappa'\,=\,\sin^2\al$.\\

\begin{figure}[!tb]
  \includegraphics[width=0.65\textwidth]{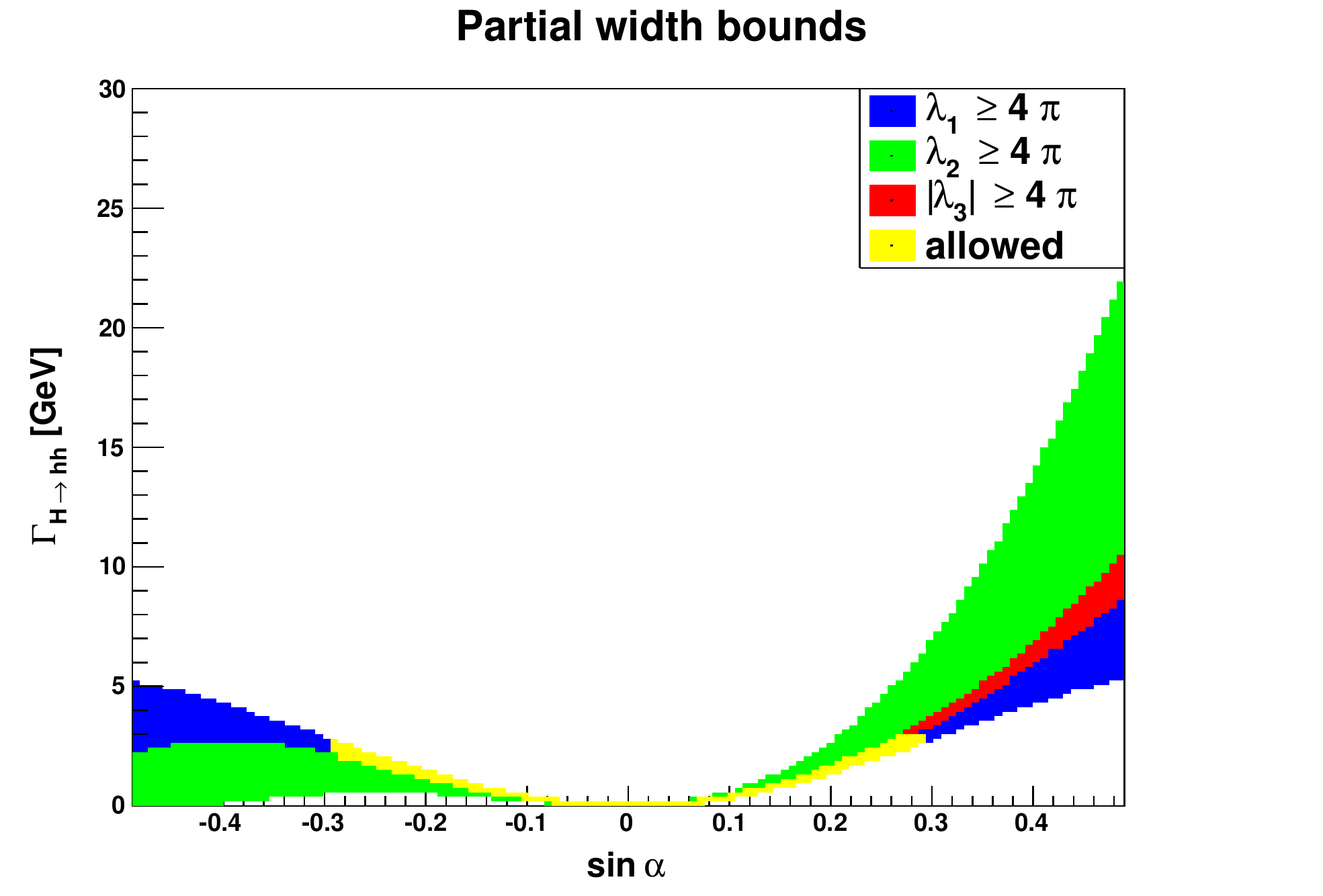}
\caption{\label{fig:gammacut} Exclusion of partial widths from perturbativity at the low scale for $|\sin\al|\,\leq\,0.49$.}
\end{figure}

For a better understanding of the effect of the constraints on the $(\Gamma,\kappa)$ parameter space, we first investigate the $H\,\rightarrow\,h\,h$ branching ratio. Figure \ref{fig:gammacut} shows the constraints in this decay width from RGE running to the scale defined by $\mu_\text{run}\,=\,2.7\,\times\,10^{10}\,\GeV$. 
We see that different regions are excluded, depending on the sign of $\sin\al$, where the biggest effects stem from perturbativity of $\lam_2$. As discussed in Section \ref{Sec:Model}, the $H\,\rightarrow\,h\,h$ squared coupling is approximately even under a sign change of $\sin\al$ for small $\tan\be$ values; if $\tan\be$ increases, it is larger for positive $\sin\al$ values. In addition, for positive (negative) $\sin\al$ values, the absolute value of the coupling decreases (increases) for decreasing $\tan\be$. Taking this into account, the exclusion bounds from perturbativity of the couplings green plotted in Figure \ref{fig:gammacut} are clear to interpret: for positive values of $\sin\al$, larger $\tan\be$ values lead to larger decay widths, and therefore large values of $\Gamma_{H\,\rightarrow\,h\,h}$ are here excluded from $\lam_2$ perturbativity, while smaller ones are affected by $\lam_1$ limits, in accordance with the limits in Figure \ref{fig:rge_run}. For negative $\sin\al$ values, the roles of $\lam_1$ and $\lam_2$ are interchanged, while smaller values of $\Gamma_{H\,\rightarrow\,h\,h}$ are achieved due to smaller absolute values of the coupling.\\

\begin{figure}[!tb]
\begin{minipage}{0.49\textwidth}
\includegraphics[width=1.1\textwidth]{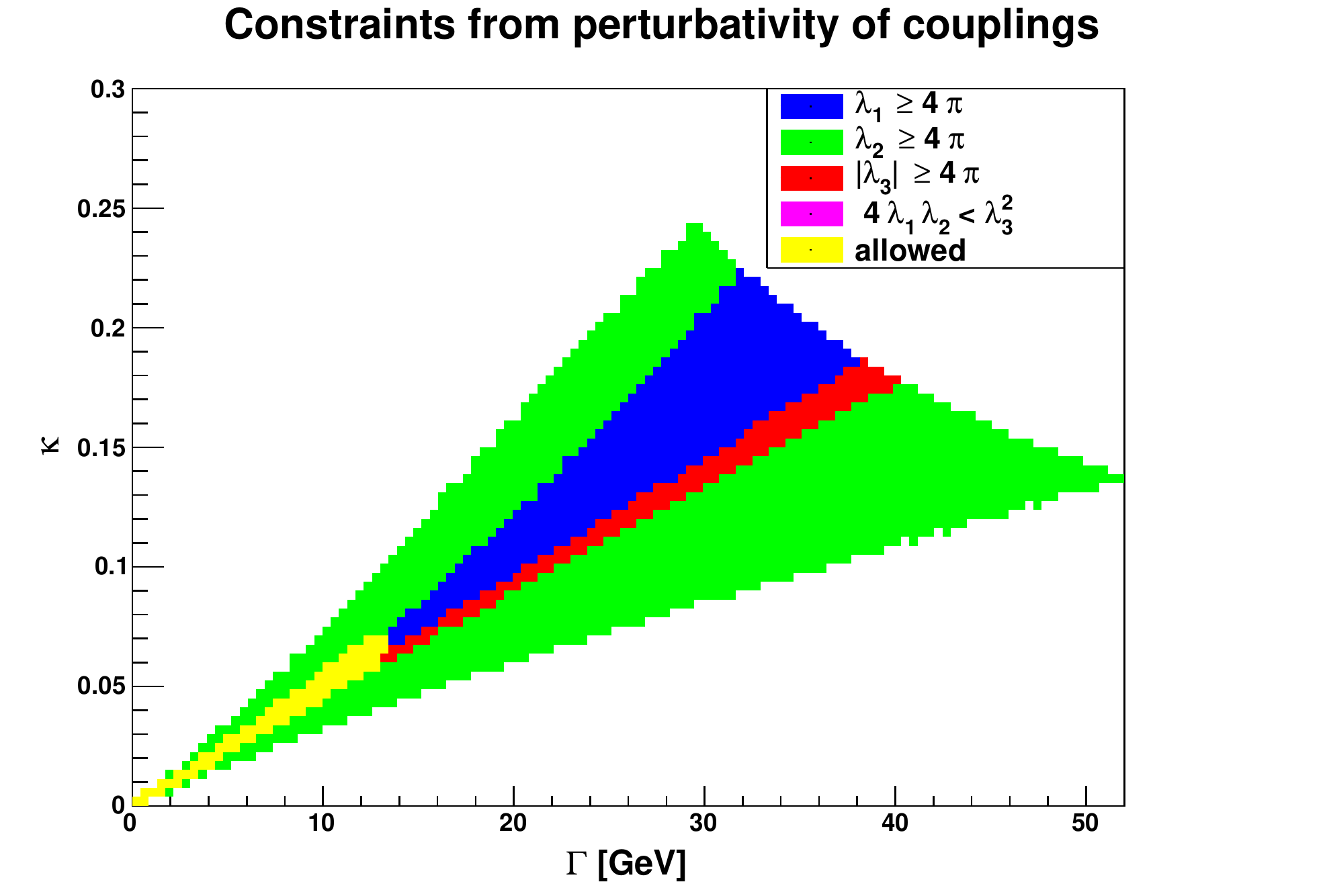}
\end{minipage}
\begin{minipage}{0.49\textwidth}
\includegraphics[width=1.1\textwidth]{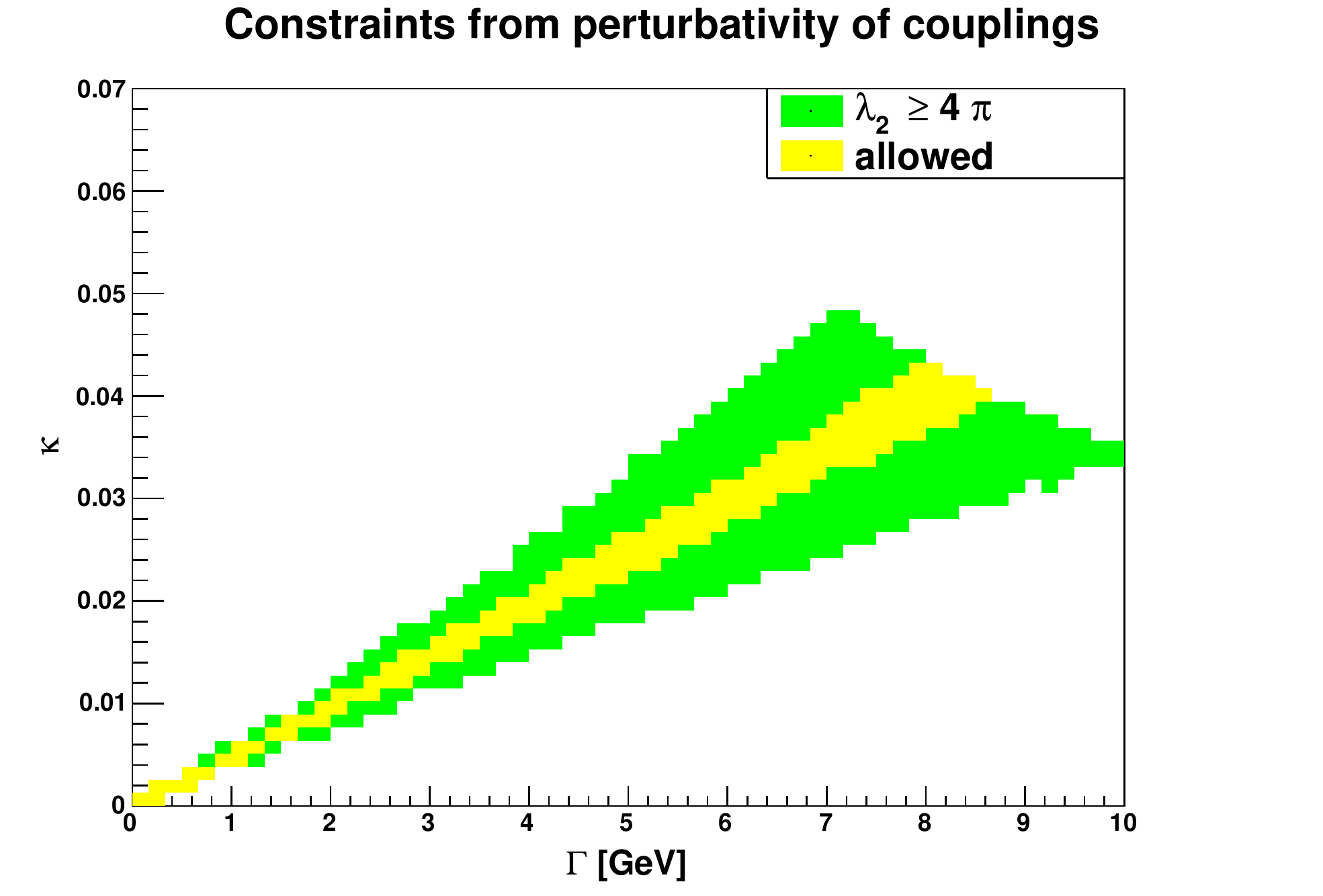}
\end{minipage}
\caption{\label{fig:gk} Constraints on the $(\Gamma,\kappa)$ plane from perturbative running of the couplings ($m_H\,=\,600\,\GeV$) to $\mu_\text{run}\,=\,2.7\,\times\,10^{10}\,\GeV$. {\sl LEFT:} Constraints for $|\sin\al|\,\leq\,0.49$; {\sl RIGHT:} Constraints when the light Higgs signal strength is taken into account.}
\end{figure}

\begin{figure}[!tb]
\begin{minipage}{0.49\textwidth}
\includegraphics[width=1.1\textwidth]{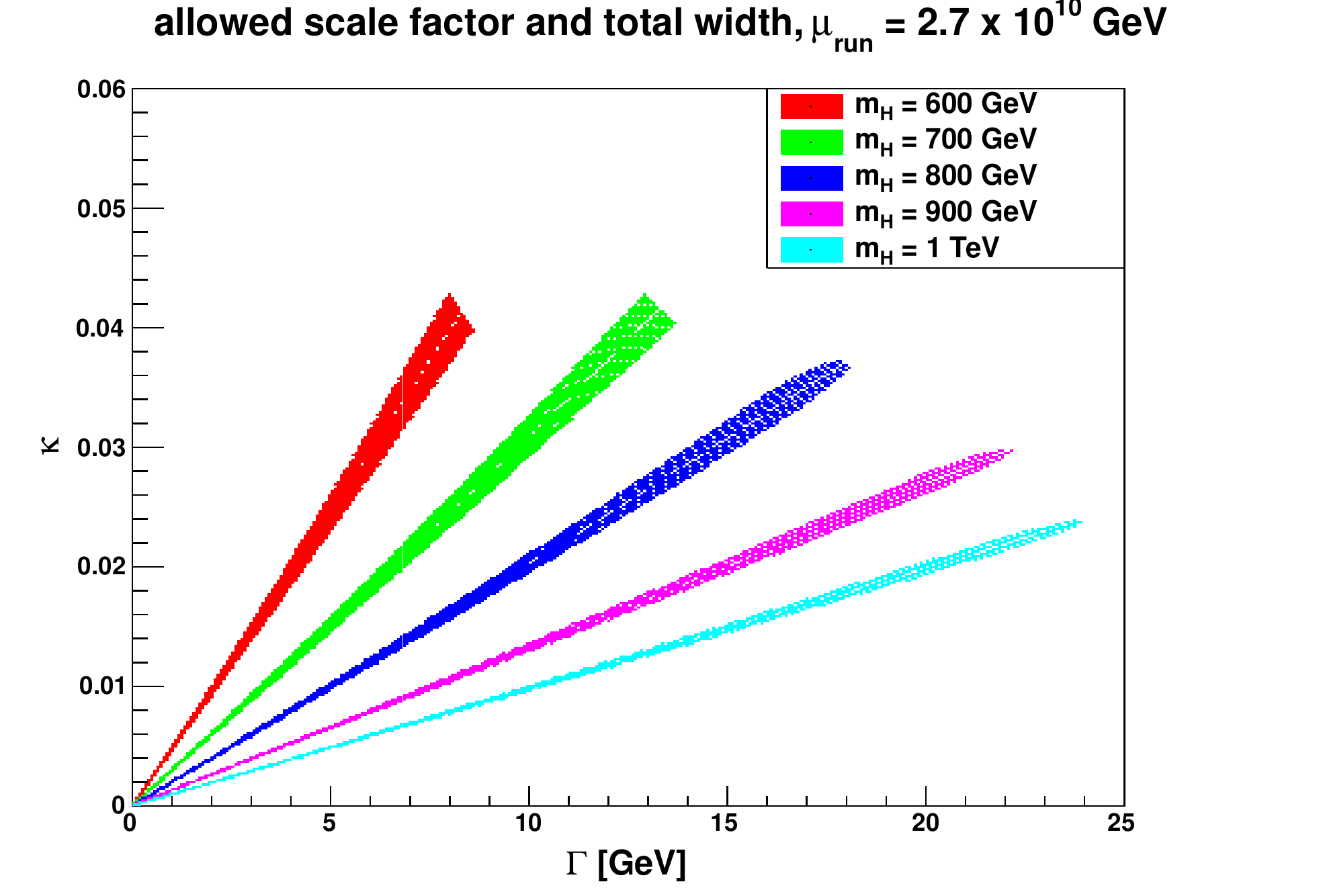}
\end{minipage}
\begin{minipage}{0.49\textwidth}
 \includegraphics[width=1.1\textwidth]{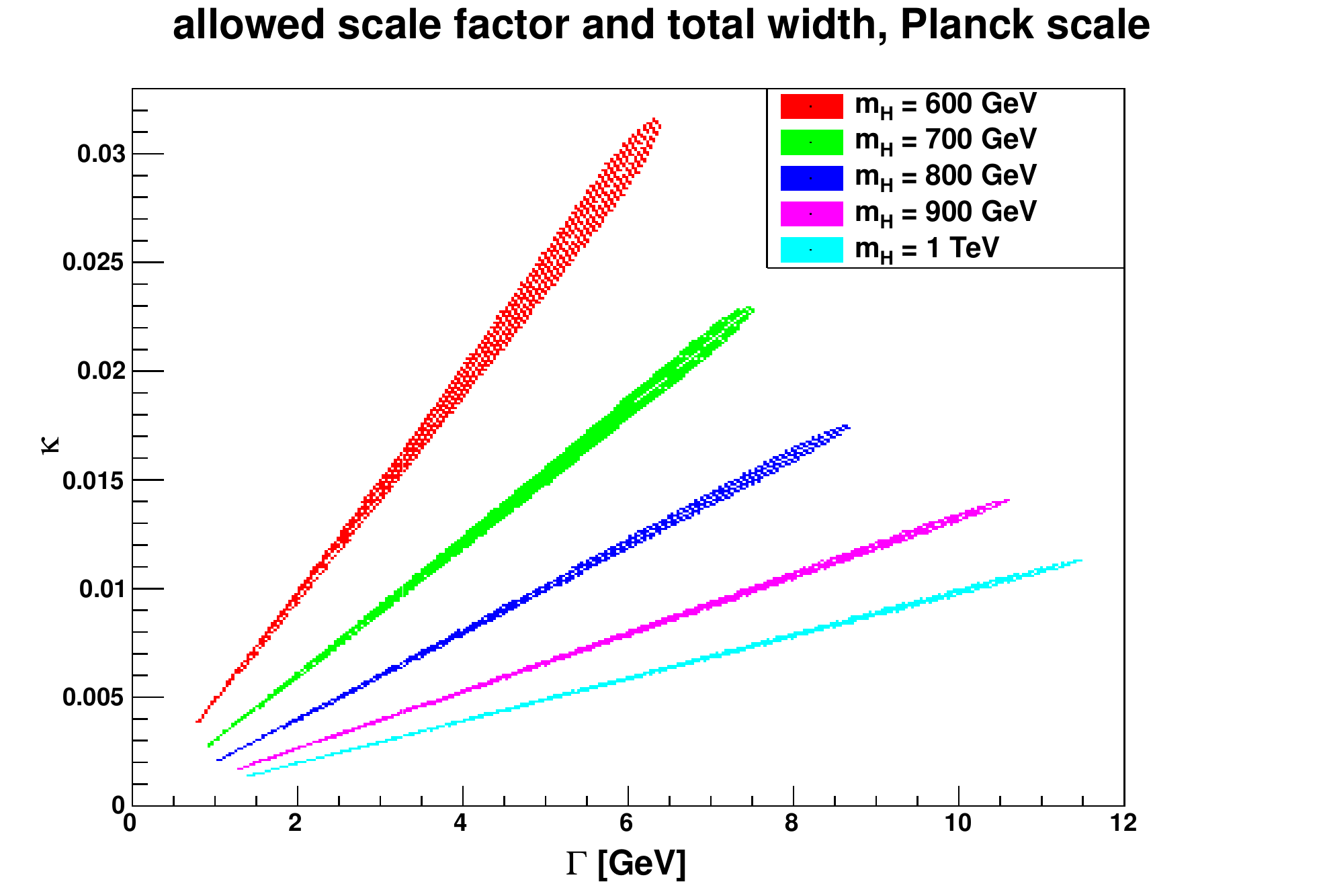}
\end{minipage}
\caption{\label{fig:gk_scales} Exclusion bounds in $(\Gamma,\, \kappa)$ plane from perturbativity, different heavy Higgs masses, for RGE running up to $\mu_\text{run}\,=\,2.7\,\times\,10^{10}\,\GeV$ {\sl (left)} as well as the Planck scale {\sl (right)}.}
\end{figure}

Figures \ref{fig:gk} and \ref{fig:gk_scales} then show how the cuts translate on the allowed parameter space in the $(\Gamma,\kappa)$ plane. The most important result is that the limitation of the angle comes with a maximally allowed total width $\Gamma_\text{tot}\,\lesssim\,14\,\GeV$ (reducing to $\,\lesssim\,6\,\GeV$ if we require perturbativity up to the Planck scale) for a Higgs mass of 600 \GeV~; this should be compared with the SM-like Higgs Boson width of $\sim\,100\,\GeV$.  For a $1\,\TeV$ Higgs, the maximal values are $25\,\GeV$ for perturbativity at the low scale and $12\,\GeV$ for perturbativity at the Planck scale, respectively. Vacuum stability cuts out lower regions of the mixing angles/ $\Gamma_{H\,\rightarrow\,h\,h}$; here, it is important to note that the {\sl minimally} allowed widths are in the sub-GeV range. The limits from this on the $(\Gamma,\,\kappa)$ plane are more pronounced for higher scales and lead to minimal values of $\Gamma_\text{min}\,=\,0.79\,\GeV\,(1.4\,\GeV),\,\kappa_\text{min}\,=\,0.004\,(0.001)$ for validity up to the Planck scale for a $600\,\GeV\,(1\,\TeV)$ heavy Higgs.\\

Finally, we remark that, if the limits from RGE are not considered, all other constraints are much less stringent; only the upper limit on $\tan\be$  from perturbative unitarity cuts out a small region in the large positive $\sin\al$/ large $\Gamma$ and large negative $\sin\al$/ low $\Gamma$ region, similarly to the constraints that we obtain from perturbativity requirements of  $\lam_2$.\\


\begin{figure}[!tb]
\begin{minipage}{0.49\textwidth}
\includegraphics[width=1.1\textwidth]{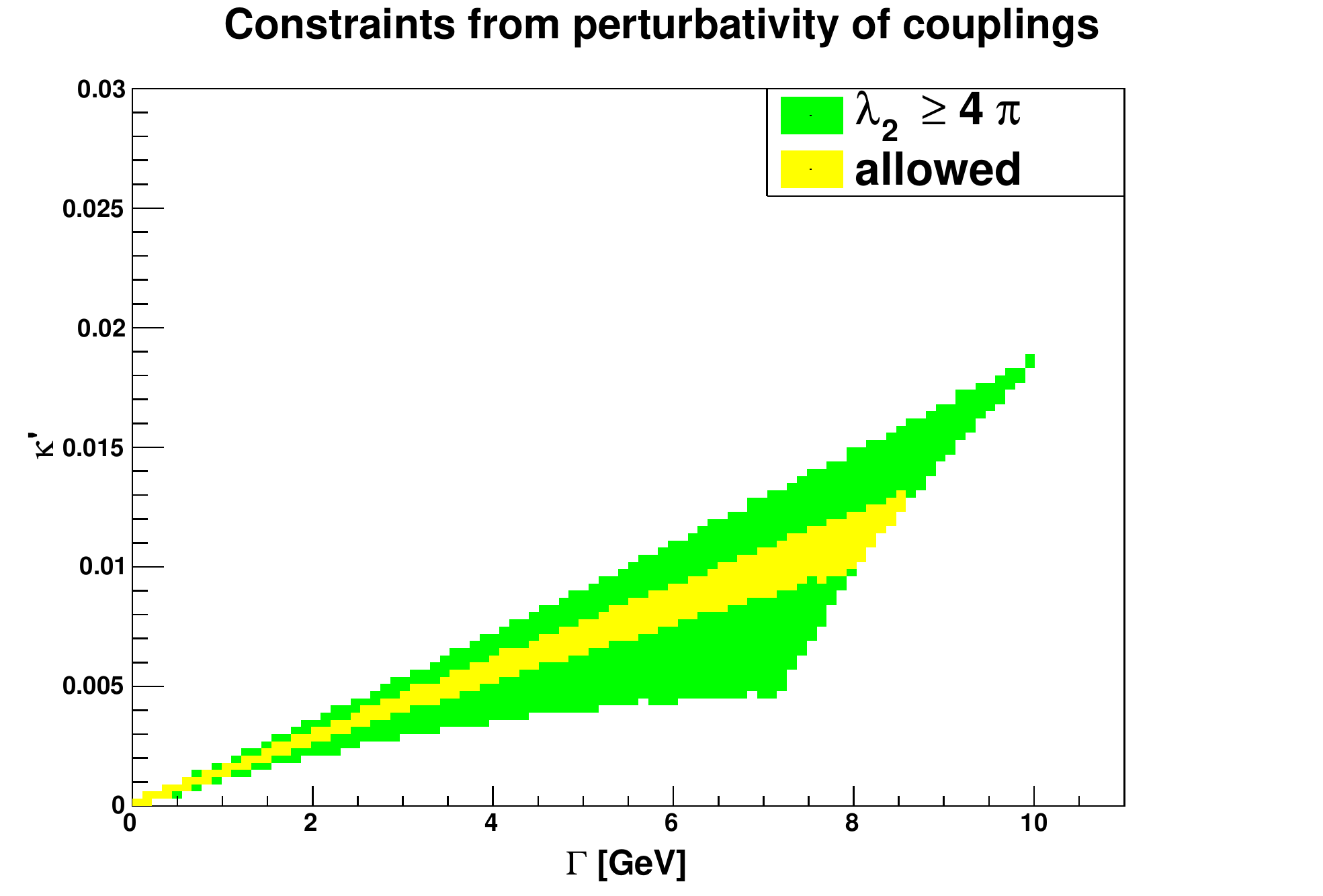}
\end{minipage}
\begin{minipage}{0.49\textwidth}
\includegraphics[width=1.1\textwidth]{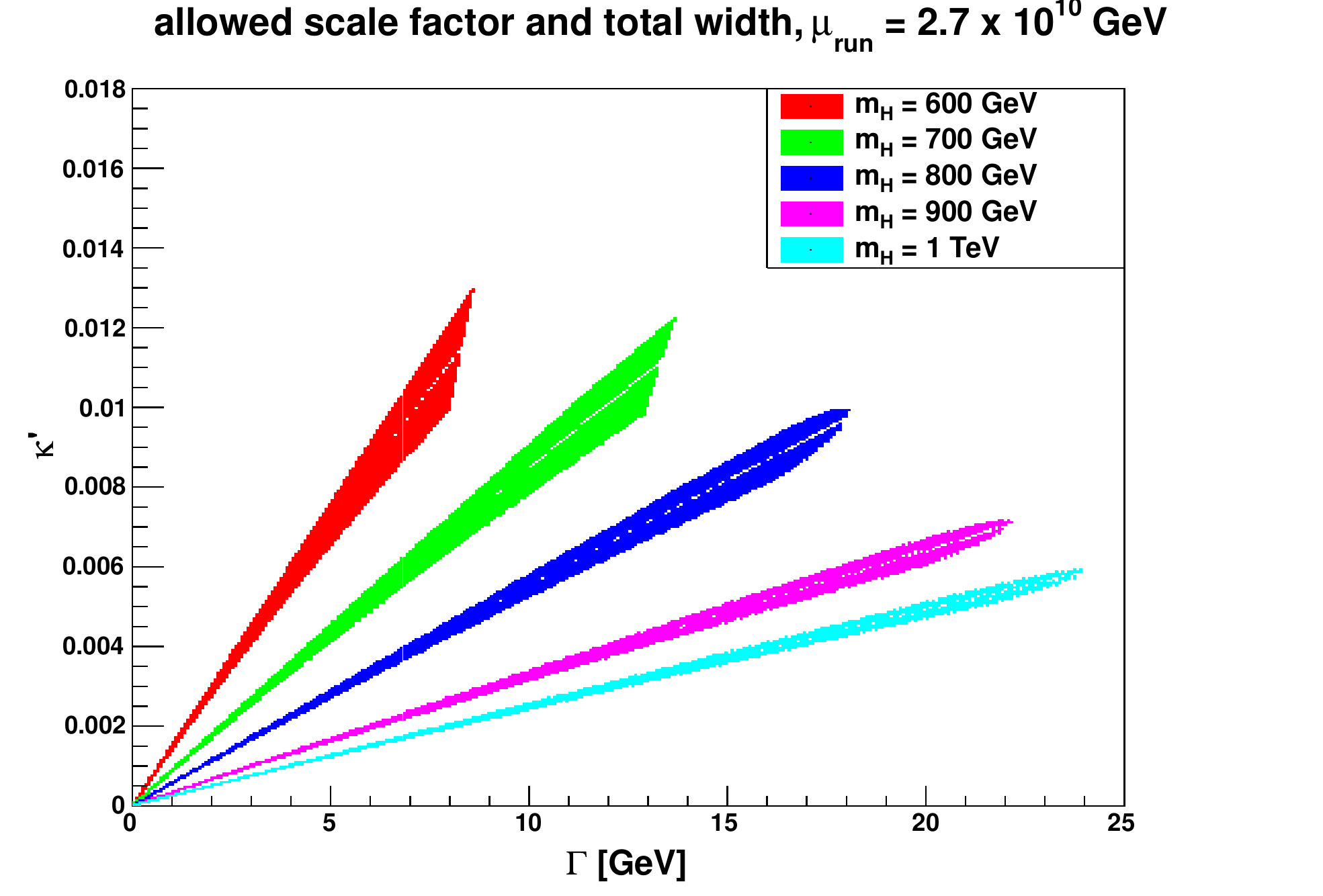}
\end{minipage}
\caption{\label{fig:gkp} Exclusion bounds in ($\Gamma,\, \kappa'$) plane from RGEs, $\mu_\text{run}\,=\,2.7\,\times\,10^{10}\,\GeV$. {\sl LEFT:} effect of perturbativity of different couplings. As before, the strongest constraints stem from the requirement of perturbativity of $\lam_2$. {\sl RIGHT:}  Results when varying $m_H$. The maximal value of $\kappa'$ is in the $\%$ range for $m_H\,=\,600\,\GeV$. With increasing $m_H$, $\kappa'_\text{max}$ decreases. }
\end{figure}

We find that the maximally allowed values for $\kappa$ are roughly  $(0.04;\,0.04;\,0.04;\,0.03;\,0.025)$ for $m_H\,=\,(600;\,700;\,800;\,900;\,1000)\,\GeV$. Concerning collider searches, the best prospect is therefore the search of a relatively light Higgs Boson at $m_H\,=\,600\,\GeV$, which would lead to $0.013\,(0.093)\,\pb$ at a $7\,(14),\,\TeV$ LHC in the gluon fusion and $2\,\times\,10^{-3}\,(0.016)\,\pb$ for the vector Boson fusion channel\footnote{Production cross sections have been taken from \cite{Dittmaier:2011ti}.}. However, also note that the widths for all masses are $\lesssim\,25\,\GeV$, which might allow for new search strategies at such masses for narrow scalar resonances\footnote{Present studies usually assume quite broad Higgses in this mass range, following the SM Higgs searches. This assumption is not consistent with our scenario.}. The maximal value of $\kappa'$, on the other hand, is 0.013 for a 600 \GeV~ Higgs mass, which would lead to a total cross section of $4\,\times\,10^{-3}\,\pb\,(0.03\,\pb)$ from gluon gluon fusion production at a 7 \TeV (14 \TeV) LHC for the  additional channel $H\,\rightarrow\,h\,h$. Other allowed values of $(\kappa',\Gamma)$ for different Higgs masses at the low $(\mu_\text{run}\,=\,2.7\,\times\,10^{10}\,\GeV)$ scale can be obtained from Figure \ref{fig:gkp}. {\cblack If, as briefly mentioned in Section \ref{sec:rgerun}, we relax the requirements of both perturbative unitarity as well as electroweak symmetry breaking at high scales, this basically opens up the parameter space for smaller positive mixing angles, effectively leading to lower minimal values of $\kappa,\,\Gamma$. The effects are negligible for the low scale; for the Planck scale, the minimal allowed width is decreased to $\sim\,0.5\,\GeV$ for nearly all masses considered here. However, as the small mixing range will be hard to detect at colliders, there is no visible impact from this on the above discussion of collider observables.} 

Finally, we discuss the allowed regions in the ($\kappa', \kappa$) plane, when all bounds are taken into account. The ratio of these two quantities is related to branching ratios for both SM-like as well as the BSM $hh$ final states:
\begin{\eqn*}
\frac{\sum \text{BR}_{H\,\rightarrow\,\text{SM}}}{\text{BR}_{H\rightarrow\,h\,h}}\,=\,\frac{\kappa}{\kappa'},
\end{\eqn*}
where $\sum \text{BR}_{H\,\rightarrow\,\text{SM}}$ now denotes the sum over all branching ratios leading to a SM-like final state. A specific branching ratio for a distinct SM-like final state $XY$ for a given heavy Higgs mass $m_H$ can then be determined via
\begin{\eqn*}
\text{BR}_{H\,\rightarrow\,XY}(m_H)\,=\,\lb\sum \text{BR}_{H\,\rightarrow\,\text{SM}}\rb\,\times\,\text{BR}^\text{SM}_{H\,\rightarrow\,XY}(m_H),
\end{\eqn*} 
where $\text{BR}^\text{SM}_{H\,\rightarrow\,XY}(m_H)$ denotes the branching ratio of a SM-like Higgs with mass $m_H$ into the final state $X\,Y$.
As $\kappa,\,\kappa'$ are indeed the parameters which can directly be observed (or constrained) at the LHC, considering the relation between these parameters provides additional useful information.
\begin{figure}[!tb]
\begin{minipage}{0.49\textwidth}
\includegraphics[width=1.1\textwidth]{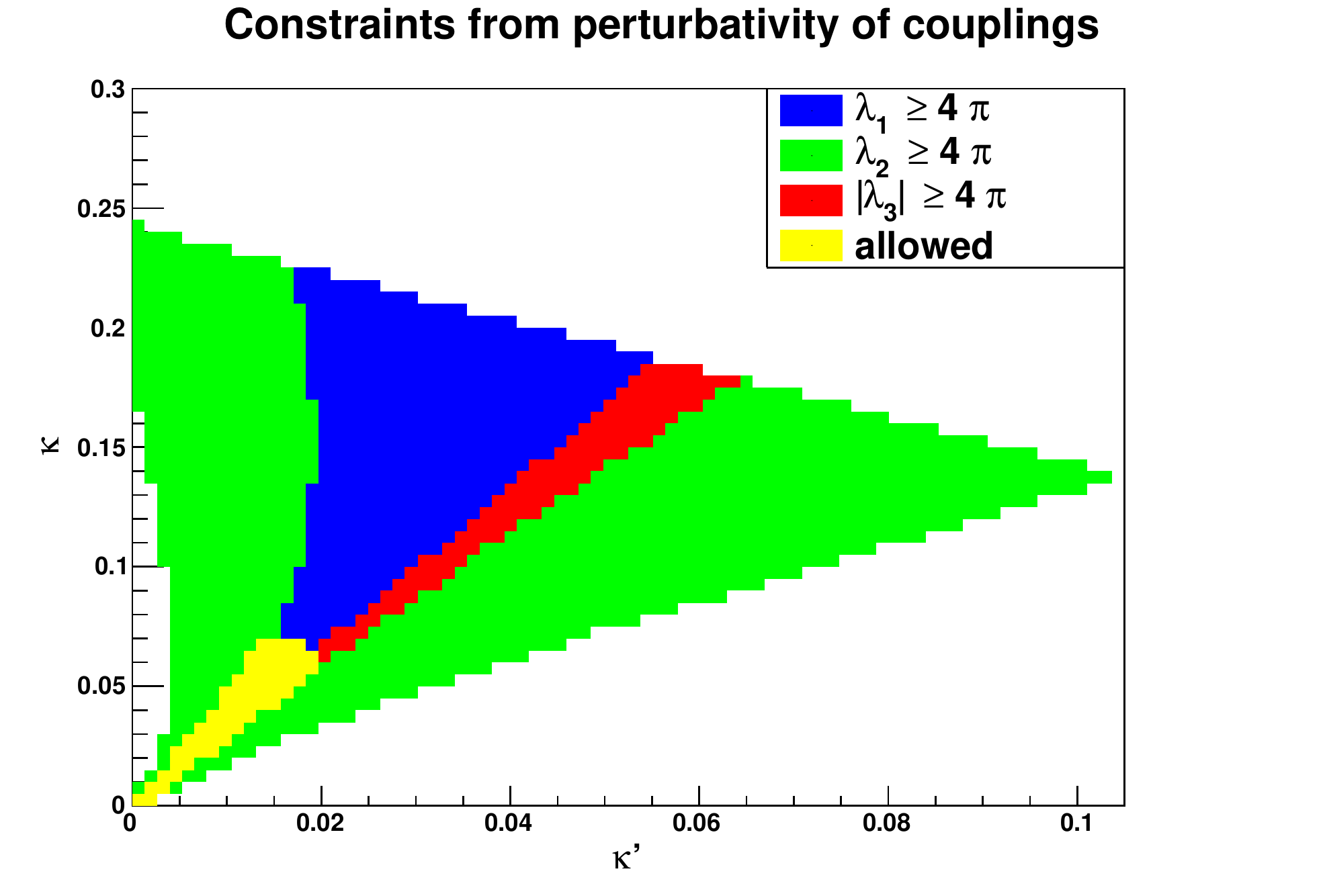}
\end{minipage}
\begin{minipage}{0.49\textwidth}
\includegraphics[width=1.1\textwidth]{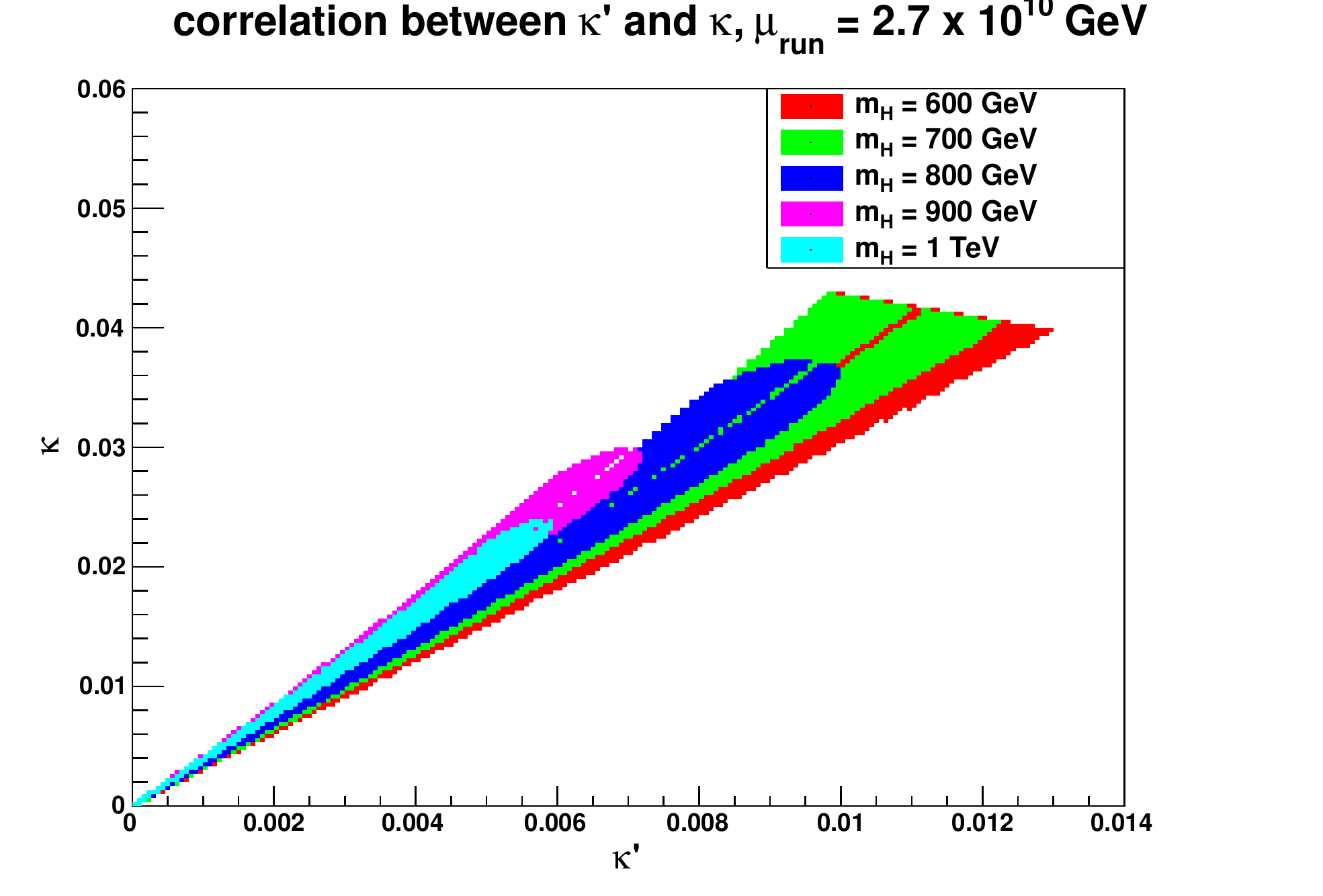}
\end{minipage}
\caption{\label{fig:kpk} Exclusion bounds in ($\kappa',\,\kappa$) plane from RGEs, $\mu_\text{run}\,=\,2.7\,\times\,10^{10}\,\GeV$. {\sl LEFT:} effect of perturbativity of different couplings, for $m_H\,=\,600\,\GeV$ and $|\sin\al|\,\leq\,0.49$. For $|\sin\al|\,\leq\,0.23$ (and therefore $\kappa+\kappa'\,\leq\,0.05$), the strongest constraints again stem from the requirement of perturbativity of $\lam_2$. {\sl RIGHT:}  Results when varying $m_H$, where now all bounds were taken into account. Constraints on one or more of the SM-like branching ratios can directly be translated into bounds on the $H\,\rightarrow\,h\,h$ branching ratio, and vice versa. }
\end{figure}
In Figure \ref{fig:kpk}, we show the results of imposing all bounds in the $\kappa',\,\kappa$ plane. Note that by definition, $\kappa+\kappa'\,=\,\sin^2\al$, which accounts for the hard cutoff visible for $m_H\,\leq\,700\,\GeV$. We see that, independent of the Higgs mass, the allowed regions all lie within a relatively narrow strip. Therefore, limits on one of these parameters can constrain the other: for example, limits on $\kappa'$ from searches in the $H\,\rightarrow\,h\,h$ channel will allow to put bounds on $\kappa$. Independent measurements of these two quantities can in contrast serve as a viability check of our model.

\subsection*{Variation of input parameters}

Finally, we comment on how a variation of SM-input parameters affects our results. The running of the SM Higgs coupling is known to be sensitive, especially\footnote{See e.g. \cite{Degrassi:2012ry}.} to the strong coupling $\al_s$ as well as the top Yukawa coupling $y_t$, so we investigate the robustness of our results under variations of these parameters at the low scale. We here consider
\begin{\eqn*}
\al_s(m_Z)\,=\,0.1184\,\pm\,0.0007,\;y_t(m_t)\,=\,0.93587\,\pm\,0.002\,\GeV,\,m_t\,=\,173.1\,\pm\,0.7\,\GeV,
\end{\eqn*}
where, following \cite{Degrassi:2012ry}, we combine the above error in the top sector to\footnote{Note that we only want to estimate the effects of deviations from the central values; a more accurate error determination should be embedded in a higher order investigation of the Higgs singlet extension.}
\begin{\eqn*}
y_t(m_t)\,=\,0.93587\,\pm\,0.006\,\GeV
\end{\eqn*}
For the results presented so far, we have used the central values above. We comment directly on the effects of values for $\kappa,\;\Gamma$, as these are the observables of main experimental interest.

\begin{itemize}
\item lowering (raising) $\al_s$, while keeping $y_t$ fixed:\\
in this case, the allowed minimal mixing angle allowed from vacuum stability is marginally increased (decreased). A changed value of $\al_s$ mainly influences the running of $y_t$, which in turn leads to a faster (slower) descent of $\lambda_1$, mainly affecting limits from the third vaccum stability condition. Changes are however in the $\%$ regime. The upper limits of $\sin\al$, which constitute the main restrictions in the $(\Gamma,\,\kappa)$ plane, are not affected.
\item keeping $\al_s$ fixed, while raising (lowering) $y_t$:\\
for a higher (lower) $y_t$, a larger (smaller) region of small mixing angles is excluded, again due to the faster (slower) descent of $\lam_1$. Maximally allowed values as well as large $\tan\be$ limits are not changed. However, using $y_t(m_t)\,=\,0.92987$ as input value shifts the breakdown of the SM-case ($\sin\al\,=\,0)$ running of the coupling by approximately an order of magnitude to $\mu_\text{run}\,\sim\,4.4\,\times\,10^{11}\,\GeV$. Limits from perturbativity of the couplings still persist.
\end{itemize}


\section{Conclusions}
\label{Sec:Conclusions}
\noindent
In this work, we have investigated the theoretical and experimental limits of the parameter space of a pure singlet extension of the SM Higgs sector without contributions from a hidden sector, where the heavy Higgs lies in a mass range of $600\,\GeV$ to $1\,\TeV$. We found that, after the light Higgs coupling strength measurements from the LHC experiments have been taken into account, additional strong limits stem from perturbativity of the couplings as well as vacuum stability, following from the $\beta$-functions of the theory. Even for a relatively low breakdown scale $\mO(10^{10}\,\GeV)$, the running of the heavy and light Higgs self-couplings {\it severely restricts} the allowed parameter space. We have translated this into observables which are currently tested by the LHC experiments, i.e. a global rescaling factor $\kappa$ for SM-like decay modes of the model, as well as total width $\Gamma$ of the new scalar. In the heavy Higgs mass range considered in this work, the light Higgs Boson signal strength restricts $\kappa$ to 0.04 for $m_H\,\lesssim\,700\,\GeV$ (at the low scale), while for higher masses additional constraints arise from the running of the couplings . Hence, the searches for such a Boson at the $7\,/8\,\TeV$ with a relatively low luminosity are surely challenging. However, on the upside we found that the total width of the new scalars is usually quite suppressed with respect to SM Higgses of such masses, with widths lying in the $1-25\,\GeV$ range (they are always $\lesssim\,0.02\,m_H$). In addition, we have introduced a second scaling parameter $\kappa'$ which parametrizes the additional decay $H\,\rightarrow\,h\,h$. We found that maximal values of this parameter are in the $\%$ range.\\

In our work, we have neglected additional contributions in the $\beta$-functions which might modify the runnings and eventually enhance the parameter space in the $\tan\be\,\gtrsim\,0.2$ region stemming from the running of $\lam_2$. These contributions, which could originate from the hidden sector,  would have to be large and negative, cancelling the rapid rising of the couplings which leads to the exclusion of experimentally interesting regions with scaling factors $\kappa$ being limited by the mixing angle alone. Scenarios with larger $\kappa'$ values which parametrizes $\Gamma_{H\,\rightarrow\,h\,h}$ are equally suppressed by the running of $\lam_2$. We plan to investigate such options and the corresponding phenomenological implications in future work.



\subsection*{Acknowledgements} 
\noindent
GMP and TR would like to thank Lorenzo Basso, Sara Bolognesi, Terrance Figy, Michael Kobel, Wolfgang Mader, and Dominik St\"ockinger for helpful discussions, as well as David Miller for useful comments regarding the manuscript. GMP is also grateful to the INFN (Sezione di Cagliari) for the hospitality and logistical support during the completion of this work. The work of GMP has been supported by the German Research Foundation DFG through Grant No.\ STO876/2-1 and by BMBF Grant No.\ 05H09ODE. In addition, the research leading to these results has partially received funding from
the
European Community's Seventh Framework Programme (FP7/2007-2013) under grant
agreement n.°290605 (PSI-FELLOW/COFUND). {\cblack Finally, we want to comment that in parallel to our work, a similar study appeared \cite{Basso:2013vla} which discusses the same Higgs sector in presence of an additional $U_{B-L}$ symmetry. We thank the author for bringing this to our attention.}


\appendix


\section{Discussion of Higgs potential and vacuum stability conditions}\label{app:hpot}
In this section, we briefly guide the reader through the steps from Eq (\ref{potential}) to Eq (\ref{bound_pot}), using the definition of the scalar fields given in Eq (\ref{unit_higgs}). 
Since the minimisation procedure is not affected by the choice of the
gauge, it is not restrictive to define the two VEVs in the following
way:
\begin{equation}\label{VEVs}
\left< H \right> \equiv
\left(
\begin{gathered}
0 \\
\frac{v}{\sqrt{2}}
\end{gathered} \right), 
\hspace{2cm}
\left< \chi \right> \equiv \frac{x}{\sqrt{2}},
\end{equation} 
with $v$ and $x$ real and non-negative.

Then, the search for extrema of $V$ is made by means of the following
differential set of equations:
\begin{equation}\label{minimisation}
\left\{
\begin{aligned}
\frac{\partial V}{\partial v}(v,x) &=& v \cdot \left( -m^2\,+\, \lambda_1 v^2 +
\frac{\lambda_3}{2}x^2 \right)=0 \\
 \frac{\partial V}{\partial x}(v,x) &=& x \cdot \left( -\mu^2\,+\, \lambda_2 x^2 +
\frac{\lambda_3}{2}v^2 \right)=0
\end{aligned}
\right.
\end{equation}

The physically interesting solutions are the ones obtained for $v$,
$x>0$:
\begin{eqnarray}\label{min_sol1}
v^2 &=& \frac{\lambda_2 m^2 - \frac{\lambda_3}{2} \mu ^2}{\lambda_1
  \lambda_2 - \frac{\lambda_3^{\phantom{o}2}}{4}}, \\
\nonumber  \\ 
\label{min_sol2}
x^2 &=& \frac{\lambda_1 \mu^2 - \frac{\lambda_3}{2} m ^2}{\lambda_1
  \lambda_2 - \frac{\lambda_3^{\phantom{o}2}}{4}}.
\end{eqnarray}

Since the denominator in equations~(\ref{min_sol1})-(\ref{min_sol2})
is always positive (assuming that the potential is well-defined), it
follows that the numerators are forced to be positive in order to
guarantee a positive-definite non-vanishing solution for $v$ and $x$.

In order to identify the extrema, we need to evaluate the Hessian
matrix:
\begin{equation}\label{hessian}
\mathcal{H}(v,x)\equiv \left(
\begin{aligned}
\frac{\partial^2 V}{\partial v^2} &\ & \frac{\partial^2 V}{\partial v
  \partial x} \\
\frac{\partial^2 V}{\partial v \partial x} &\ & \frac{\partial^2
  V}{\partial x^2}
\end{aligned}
\right) = \left(
\begin{aligned}
2 \lambda_1 v^2 &\ & \lambda_3 v x \\
\lambda_3 v x &\ & 2 \lambda_2 x^2
\end{aligned}
\right) .
\end{equation}

From this equation, it is straightforward to verify that the solutions
are minima if and only if equations~(\ref{bound_pot}) are
satisfied.

To compute the scalar masses, one must expand the potential in
equation~(\ref{potential}) around the minima found in
equations~(\ref{min_sol1})-(\ref{min_sol2}). Then, Eqns. (\ref{mh1}),(\ref{mh2})  follow immediately.

\section{Perturbative unitarity} 

In this section we want to briefly explain the techniques that
we used in order to obtain bounds from perturbative unitarity, firstly described in detail
by \cite{Lee:1977eg}. Evaluating the tree-level scattering
amplitude of longitudinally polarised vector bosons one finds that the
latter grows with the energy of the process, eventually violating
unitarity, unless one includes some other (model dependent)
interactions. According to the equivalence theorem, the amplitude of any process with external longitudinal vector
bosons $V_L$ ($V = W^\pm,Z$) can be substituted each one of them with the related Goldstone bosons $v
= w^\pm,z$ \cite{Chanowitz:1985hj} for energies much larger than the vector Boson mass. 

Given a tree-level scattering amplitude between two spin-$0$ particles,
$M(s,\theta)$, where $\theta$ is the scattering (polar) angle, 
we know that the partial wave amplitude with angular
momentum $J$ is given by
\begin{eqnarray}\label{integral}
a_J = \frac{1}{32\pi} \int_{-1}^{1} d(\cos{\theta}) P_J(\cos{\theta})
M(s,\theta),
\end{eqnarray}
where $P_J$ are Legendre polynomials. It has been proven
(see \cite{Luscher:1988gc}) that, in order to preserve unitarity, each
partial wave must be bounded by the condition
\begin{eqnarray}\label{condition_app}
|\textrm{Re}(a_J(s))|\leq \frac{1}{2}.
\end{eqnarray}

As discussed previously, in the high energy limit, $\sqrt
s\rightarrow \infty$, only the $a_0$ partial wave amplitude does not
vanish; therefore, we here present all $a_0$'s in the high energy
and small gauge coupling limit (i.e. $\sqrt{s}\to \infty$ and $e\to 0$, respectively):
\begin{eqnarray}\label{a0zz}
a_0(zz\rightarrow zz)
&=& 
\frac{3 (m_h^2+m_H^2+(m_h^2-m_H^2) \cos\,(2\,\al))}{64 \pi  v^2} \\
\label{a0zzww}
a_0(zz\rightarrow w^+w^-)
&=& 
\frac{m_h^2+m_H^2+(m_h^2-m_H^2) \cos\,(2\,\al)}{32 \sqrt{2}\pi  v^2} \\
\label{a0zzhh}
a_0(zz\rightarrow hh)
&=&  
\frac{\cos\al}{128 \pi  v^2 x} \left((3 m_h^2+m_H^2) x \cos\al + \right. \nonumber \\
&+& \left. (m_h^2-m_H^2) \left(x \cos(3\,\al)-4 v \sin^3\al\right)\right) \\
\label{a0zzhH}
a_0(zz\rightarrow hH)
&=& 
\frac{\cos\al \sin\al ((m_h^2+m_H^2) x+(m_h^2-m_H^2) (x \cos\,(2\,\al)+v \sin(2\al)))}{32 \sqrt{2}\pi  v^2 x} \\
\label{a0zzHH}
a_0(zz\rightarrow HH)
&=&  
\frac{\sin\al}{64 \pi  v^2 x}\left(2 (-m_h^2+m_H^2) v \cos^3\al+ \right. \nonumber \\
&+& \left. x (m_h^2+m_H^2+(m_h^2-m_H^2) \cos\,(2\,\al)) \sin\al\right)
\end{eqnarray}

\begin{eqnarray}\label{a0ww}
 a_0(w^+w^-\rightarrow w^+w^-)
&=&
\frac{m_h^2+m_H^2+(m_h^2-m_H^2) \cos\,(2\,\al)}{16 \pi  v^2} \\
\label{a0wwhh}
a_0(w^+w^-\rightarrow hh)
&=& 
\frac{\cos\al}{64 \sqrt{2}\pi  v^2 x}\left((3 m_h^2+m_H^2) x \cos\al+\right. \nonumber \\
&+&\left. (m_h^2-m_H^2) \left(x \cos\,(3\,\al)-4 v \sin^3\al\right)\right) \\
\label{a0wwhH}
a_0(w^+w^-\rightarrow hH)
&=&
\frac{\cos\al \sin\al ((m_h^2+m_H^2) x+(m_h^2-m_H^2) (x \cos\,(2\,\al)+v \sin(2\al)))}{32 \pi  v^2 x} \\
\label{a0wwHH}
a_0(w^+w^-\rightarrow HH)
&=&
\frac{\sin\al}{32 \sqrt{2}\pi  v^2 x} \left(2 (-m_h^2+m_H^2) v \cos^3\al+\right. \nonumber \\
&+&\left. x (m_h^2+m_H^2+(m_h^2-m_H^2) \cos\,(2\,\al)) \sin\al\right)
\end{eqnarray}

\begin{eqnarray}\label{a0hh}
a_0(hh\rightarrow hh)
&=& 
\frac{1}{1024 \pi  v^2 x^2}\left(6 (5 m_h^2+m_H^2) \left(v^2+x^2\right)+\right. \nonumber \\
&-&3 (15 m_h^2+m_H^2) (v-x) (v+x) \cos\,(2\,\al)+ \nonumber \\
&+& 6 (3 m_h^2-m_H^2) \left(v^2+x^2\right) \cos(4\,\al)+\nonumber \\
&-&\left. 3 (m_h^2-m_H^2) \left((v-x) (v+x) \cos(6\,\al)+8 v x \sin^3(2\al)\right)\right) \\
\label{a0hhhH}
a_0(hh\rightarrow hH)
&=&
\frac{3 \cos\al \sin\al}{64 \sqrt{2}\pi  v^2 x^2}(x \cos\al+v \sin\al) ((3 m_h^2+m_H^2) x \cos\al+\nonumber \\
&+&(m_h^2-m_H^2) x \cos\,(3\,\al)-(3 m_h^2+m_H^2) v \sin\al+\nonumber \\
&+&(m_h^2-m_H^2) v \sin(3\al))\\
\label{a0hhHH}
a_0(hh\rightarrow HH)
&=& 
\frac{\sin(2\al)}{512 \pi  v^2 x^2} \left(6 (-m_h^2+m_H^2) v x \cos(4\,\al)+\right. \nonumber \\
&+&6 (m_h^2+m_H^2) \left(v^2+x^2\right) \sin(2\al)+ \nonumber \\
&-&\left. (m_h^2-m_H^2) (2 v x+3 (v-x) (v+x) \sin(4\al))\right)
\end{eqnarray}

\begin{eqnarray}\label{a0hH}
a_0(hH\rightarrow hH)
&=&
\frac{\sin(2\,\al)}{256 \pi  v^2 x^2} \left(6 (-m_h^2+m_H^2) v x \cos(4\,\al)+\right. \nonumber \\
&+& 6 (m_h^2+m_H^2) \left(v^2+x^2\right) \sin(2\,\al)+\nonumber \\
&-& \left. (m_h^2-m_H^2) (2 v x+3 (v-x) (v+x) \sin(4\al))\right) \\
\label{a0hHHH}
a_0(hH\rightarrow HH)
&=&
-\frac{3 \cos\al \sin\al}{64 \sqrt{2}\pi  v^2 x^2}(v \cos\al-x \sin\al) ((m_h^2+3 m_H^2) v \cos\al+\nonumber \\
&+&(-m_h^2+m_H^2) v \cos\,(3\,\al)+\nonumber \\
&+&2 x (m_h^2+m_H^2+(m_h^2-m_H^2) \cos\,(2\,\al)) \sin\al)
\end{eqnarray}

\begin{eqnarray}\label{a0HH}
a_0(HH\rightarrow HH)
&=&
\frac{1}{1024 \pi  v^2 x^2}\left(6 (m_h^2+5 m_H^2) \left(v^2+x^2\right)+\right. \nonumber \\
&+&3 (m_h^2+15 m_H^2) (v-x) (v+x) \cos\,(2\,\al)+\nonumber \\
&-& 6 (m_h^2-3 m_H^2) \left(v^2+x^2\right) \cos(4\,\al)+\nonumber \\
&-&\left. 3 (m_h^2-m_H^2) \left((v-x) (v+x) \cos(6\,\al)+8 v x \sin^3(2\,\al)\right)\right)
\end{eqnarray}

\label{appe:b}

\section{Analytic solution for SM gauge coupling RGEs}
In the SM, all one-loop RGEs for gauge couplings are of the form
\begin{\eqn*}
\frac{dx}{dt}\,=\,a\,\,x^2.
\end{\eqn*}
The exact analytic solution for this equation is given by
\begin{\eqn}\label{eq:rge_sol}
x\lb t \rb\,=\,\frac{x\lb t\,=\,t_0 \rb}{1-a\,x(t=t_0)\,\lb t-t_0 \rb},
\end{\eqn}
where for $t\,=\,\log\lb \frac{\lambda^2}{\lambda_\text{ref}^2} \rb$ we have
\begin{\eqn*}
t-t_0\,=\,2\,\log\lb \frac{\lambda}{\lambda_0} \rb.
\end{\eqn*}
For positive values of $a$, the coupling reaches the Landau pole when the denominator in Eq. (\ref{eq:rge_sol}) goes to 0; for negative values, $x\,\rightarrow\,0$ for $t\,\rightarrow\,\infty$.\\

Now we turn to the Yukawa coupling terms. This generic equation has the form
\begin{\eqn*}
\frac{dx}{dt}\,=\,a\,x+b\,x^3
\end{\eqn*}
with the solution
\begin{\eqn*}
x\lb t \rb\,=\,\frac{\sqrt{a\,C'(t_0)}\,e^{a\,(t-t_0)}}{\sqrt{1-b\,e^{2\,a(t-t_0)}\,C'(t_0)}},
\end{\eqn*}
with $C'(t_0)\,=\,\frac{x^2_0}{a+b\,x_0^2}$ where $x(t=t_0)\,\equiv\,x_0$ defines the initial value. In case of the top Yukawa coupling, we have
\begin{eqnarray*}
16\,\pi^2\,a&=&-4\,g_s^2-\frac{9}{8}g^2-\frac{17}{24}g'^2;\;16\,\pi^2\,b\,=\,\frac{9}{4}.
\end{eqnarray*}
However, taking the time dependence of the SM gauge couplings into account, the above solution needs to be modified such that $a\,(t-t_0)$ is replaced by $\int ^t_{t_0} a(t')\,dt'$. Although this is still feasible at one loop, we chose to solve the RGE of the top Yukawa coupling numerically\footnote{See \cite{Chishtie:2007qp} for an all-analytic solution to the first order RGEs.}.


\newpage
\section*{Bibliography}
\bibliography{bibl}

\begin{thebibliography}{10}

\bibitem{atlres}
Georges Aad et~al.
\newblock {Observation of a new particle in the search for the Standard Model
  Higgs boson with the ATLAS detector at the LHC}.
\newblock {\em Phys.Lett.}, B716:1--29, 2012.

\bibitem{cmsres}
Serguei Chatrchyan et~al.
\newblock {Observation of a new boson at a mass of 125 GeV with the CMS
  experiment at the LHC}.
\newblock {\em Phys.Lett.}, B716:30--61, 2012.

\bibitem{Higgs:1964ia}
Peter~W. Higgs.
\newblock {Broken symmetries, massless particles and gauge fields}.
\newblock {\em Phys.Lett.}, 12:132--133, 1964.

\bibitem{Higgs:1964pj}
Peter~W. Higgs.
\newblock {Broken Symmetries and the Masses of Gauge Bosons}.
\newblock {\em Phys.Rev.Lett.}, 13:508--509, 1964.

\bibitem{Englert:1964et}
F.~Englert and R.~Brout.
\newblock {Broken Symmetry and the Mass of Gauge Vector Mesons}.
\newblock {\em Phys.Rev.Lett.}, 13:321--323, 1964.

\bibitem{Guralnik:1964eu}
G.S. Guralnik, C.R. Hagen, and T.W.B. Kibble.
\newblock {Global Conservation Laws and Massless Particles}.
\newblock {\em Phys.Rev.Lett.}, 13:585--587, 1964.

\bibitem{Kibble:1967sv}
T.W.B. Kibble.
\newblock {Symmetry breaking in nonAbelian gauge theories}.
\newblock {\em Phys.Rev.}, 155:1554--1561, 1967.

\bibitem{Schabinger:2005ei}
Robert Schabinger and James~D. Wells.
\newblock {A Minimal spontaneously broken hidden sector and its impact on Higgs
  boson physics at the large hadron collider}.
\newblock {\em Phys.Rev.}, D72:093007, 2005.

\bibitem{Patt:2006fw}
Brian Patt and Frank Wilczek.
\newblock {Higgs-field portal into hidden sectors}.
\newblock 2006.

\bibitem{Barger:2007im}
Vernon Barger, Paul Langacker, Mathew McCaskey, Michael~J. Ramsey-Musolf, and
  Gabe Shaughnessy.
\newblock {LHC Phenomenology of an Extended Standard Model with a Real Scalar
  Singlet}.
\newblock {\em Phys.Rev.}, D77:035005, 2008.

\bibitem{Bhattacharyya:2007pb}
Gautam Bhattacharyya, Gustavo~C. Branco, and S.~Nandi.
\newblock {Universal Doublet-Singlet Higgs Couplings and phenomenology at the
  CERN Large Hadron Collider}.
\newblock {\em Phys.Rev.}, D77:117701, 2008.

\bibitem{Dawson:2009yx}
Sally Dawson and Wenbin Yan.
\newblock {Hiding the Higgs Boson with Multiple Scalars}.
\newblock {\em Phys.Rev.}, D79:095002, 2009.

\bibitem{Bock:2010nz}
Sebastian Bock, Remi Lafaye, Tilman Plehn, Michael Rauch, Dirk Zerwas, et~al.
\newblock {Measuring Hidden Higgs and Strongly-Interacting Higgs Scenarios}.
\newblock {\em Phys.Lett.}, B694:44--53, 2010.

\bibitem{Fox:2011qc}
Patrick~J. Fox, David Tucker-Smith, and Neal Weiner.
\newblock {Higgs friends and counterfeits at hadron colliders}.
\newblock {\em JHEP}, 1106:127, 2011.

\bibitem{Englert:2011yb}
Christoph Englert, Tilman Plehn, Dirk Zerwas, and Peter~M. Zerwas.
\newblock {Exploring the Higgs portal}.
\newblock {\em Phys.Lett.}, B703:298--305, 2011.

\bibitem{Englert:2011us}
Christoph Englert, Joerg Jaeckel, Emanuele Re, and Michael Spannowsky.
\newblock {Evasive Higgs Maneuvers at the LHC}.
\newblock {\em Phys.Rev.}, D85:035008, 2012.

\bibitem{Batell:2011pz}
Brian Batell, Stefania Gori, and Lian-Tao Wang.
\newblock {Exploring the Higgs Portal with 10/fb at the LHC}.
\newblock {\em JHEP}, 1206:172, 2012.

\bibitem{Englert:2011aa}
Christoph Englert, Tilman Plehn, Michael Rauch, Dirk Zerwas, and Peter~M.
  Zerwas.
\newblock {LHC: Standard Higgs and Hidden Higgs}.
\newblock {\em Phys.Lett.}, B707:512--516, 2012.

\bibitem{Gupta:2011gd}
Rick~S. Gupta and James~D. Wells.
\newblock {Higgs boson search significance deformations due to mixed-in
  scalars}.
\newblock {\em Phys.Lett.}, B710:154--158, 2012.

\bibitem{Dolan:2012ac}
Matthew~J. Dolan, Christoph Englert, and Michael Spannowsky.
\newblock {New Physics in LHC Higgs boson pair production}.
\newblock {\em Phys.Rev.}, D87:055002, 2013.

\bibitem{Bertolini:2012gu}
Daniele Bertolini and Matthew McCullough.
\newblock {The Social Higgs}.
\newblock {\em JHEP}, 1212:118, 2012.

\bibitem{Batell:2012mj}
Brian Batell, David McKeen, and Maxim Pospelov.
\newblock {Singlet Neighbors of the Higgs Boson}.
\newblock {\em JHEP}, 1210:104, 2012.

\bibitem{hhwg}
cf eg C. Grojean, {\sl BSM Higgs Benchmark Scenarios}, 7th meeting of the LHC
  Higgs Cross Section Working Group, Dec 2012,
  https://indico.cern.ch/conferenceOtherViews.py?view=standard$\&$confId=20960%
5.

\bibitem{hhwg2}
cf eg S. Bolognesi, {\sl Introduction: heavy "Higgs" status and BSM
  benchmarks}, BSM Heavy Higgs meeting, April 2013,
  https://indico.cern.ch/conferenceOtherViews.py?view=standard$\&$confId=24875%
1.

\bibitem{Heinemeyer:2013tqa}
S.~Heinemeyer et~al.
\newblock {Handbook of LHC Higgs Cross Sections: 3. Higgs Properties}.
\newblock 2013.

\bibitem{Basso:2010jm}
Lorenzo Basso, Stefano Moretti, and Giovanni~Marco Pruna.
\newblock {A Renormalisation Group Equation Study of the Scalar Sector of the
  Minimal B-L Extension of the Standard Model}.
\newblock {\em Phys.Rev.}, D82:055018, 2010.

\bibitem{Strassler:2006im}
Matthew~J. Strassler and Kathryn~M. Zurek.
\newblock {Echoes of a hidden valley at hadron colliders}.
\newblock {\em Phys.Lett.}, B651:374--379, 2007.

\bibitem{Strassler:2006ri}
Matthew~J. Strassler and Kathryn~M. Zurek.
\newblock {Discovering the Higgs through highly-displaced vertices}.
\newblock {\em Phys.Lett.}, B661:263--267, 2008.

\bibitem{Bowen:2007ia}
Matthew Bowen, Yanou Cui, and James~D. Wells.
\newblock {Narrow trans-TeV Higgs bosons and H $\rightarrow$ hh decays: Two LHC
  search paths for a hidden sector Higgs boson}.
\newblock {\em JHEP}, 0703:036, 2007.

\bibitem{ATLAS-CONF-2013-012}
Measurements of the properties of the Higgs-like Boson in the two photon decay
  channel with the ATLAS detector using 25 $\mathrm{fb}^{-1}$ of proton-proton
  collision data.
\newblock Technical Report ATLAS-CONF-2013-012, CERN, Geneva, Mar 2013.

\bibitem{CMS-PAS-HIG-13-001}
Updated measurements of the Higgs Boson at 125 GeV in the two photon decay
  channel.
\newblock Technical Report CMS-PAS-HIG-13-001, CERN, Geneva, 2013.

\bibitem{Lee:1977eg}
Benjamin~W. Lee, C.~Quigg, and H.B. Thacker.
\newblock {Weak Interactions at Very High-Energies: The Role of the Higgs Boson
  Mass}.
\newblock {\em Phys.Rev.}, D16:1519, 1977.

\bibitem{Luscher:1988gc}
M.~Luscher and P.~Weisz.
\newblock {Is There a Strong Interaction Sector in the Standard Lattice Higgs
  Model?}
\newblock {\em Phys.Lett.}, B212:472, 1988.

\bibitem{Chanowitz:1985hj}
Michael~S. Chanowitz and Mary~K. Gaillard.
\newblock {The TeV Physics of Strongly Interacting W's and Z's}.
\newblock {\em Nucl.Phys.}, B261:379, 1985.

\bibitem{Basso:2010jt}
L.~Basso, A.~Belyaev, S.~Moretti, and G.M. Pruna.
\newblock {Tree Level Unitarity Bounds for the Minimal B-L Model}.
\newblock {\em Phys.Rev.}, D81:095018, 2010.

\bibitem{Peskin:1991sw}
Michael~E. Peskin and Tatsu Takeuchi.
\newblock {Estimation of oblique electroweak corrections}.
\newblock {\em Phys.Rev.}, D46:381--409, 1992.

\bibitem{Hagiwara:1994pw}
Kaoru Hagiwara, S.~Matsumoto, D.~Haidt, and C.S. Kim.
\newblock {A Novel approach to confront electroweak data and theory}.
\newblock {\em Z.Phys.}, C64:559--620, 1994.

\bibitem{Espinosa:2012im}
J.R. Espinosa, C.~Grojean, M.~Muhlleitner, and M.~Trott.
\newblock {First Glimpses at Higgs' face}.
\newblock {\em JHEP}, 1212:045, 2012.

\bibitem{Beringer:1900zz}
J.~Beringer et~al.
\newblock {Review of Particle Physics (RPP)}.
\newblock {\em Phys.Rev.}, D86:010001, 2012.

\bibitem{Gupta:2012mi}
Rick~S. Gupta, Heidi Rzehak, and James~D. Wells.
\newblock {How well do we need to measure Higgs boson couplings?}
\newblock {\em Phys.Rev.}, D86:095001, 2012.

\bibitem{ATLAS-CONF-2013-034}
Combined coupling measurements of the Higgs-like Boson with the ATLAS detector
  using up to 25 fb$^{-1}$ of proton-proton collision data.
\newblock Technical Report ATLAS-CONF-2013-034, CERN, Geneva, Mar 2013.

\bibitem{CMS-PAS-HIG-13-005}
{Combination of standard model Higgs Boson searches and measurements of the
  properties of the new Boson with a mass near 125 GeV}.
\newblock Technical Report CMS-PAS-HIG-13-005, CERN, Geneva, 2013.

\bibitem{Lebedev:2012zw}
Oleg Lebedev.
\newblock {On Stability of the Electroweak Vacuum and the Higgs Portal}.
\newblock {\em Eur.Phys.J.}, C72:2058, 2012.

\bibitem{Belanger:2012zr}
G.~Belanger, K.~Kannike, A.~Pukhov, and M.~Raidal.
\newblock {$Z_3$ Scalar Singlet Dark Matter}.
\newblock {\em JCAP}, 1301:022, 2013.

\bibitem{Degrassi:2012ry}
Giuseppe Degrassi, Stefano Di~Vita, Joan Elias-Miro, Jose~R. Espinosa, Gian~F.
  Giudice, et~al.
\newblock {Higgs mass and vacuum stability in the Standard Model at NNLO}.
\newblock {\em JHEP}, 1208:098, 2012.

\bibitem{Gunion:1989we}
John~F. Gunion, Howard~E. Haber, Gordon~L. Kane, and Sally Dawson.
\newblock {THE HIGGS HUNTER'S GUIDE}.
\newblock {\em Front.Phys.}, 80:1--448, 2000.

\bibitem{EliasMiro:2012ay}
Joan Elias-Miro, Jose~R. Espinosa, Gian~F. Giudice, Hyun~Min Lee, and
  Alessandro Strumia.
\newblock {Stabilization of the Electroweak Vacuum by a Scalar Threshold
  Effect}.
\newblock {\em JHEP}, 1206:031, 2012.

\bibitem{Dittmaier:2011ti}
S.~Dittmaier et~al.
\newblock {Handbook of LHC Higgs Cross Sections: 1. Inclusive Observables}.
\newblock 2011.

\bibitem{Basso:2013vla}
Lorenzo Basso.
\newblock {Minimal Z' models and the 125 GeV Higgs boson}.
\newblock {\em Phys.Lett.}, B725:322--326, 2013.

\bibitem{Chishtie:2007qp}
F.A. Chishtie, D.G.C. McKeon, T.G. Steele, and I.~Zakout.
\newblock {Exact One Loop Running Couplings in the Standard Model}.
\newblock {\em Can.J.Phys.}, 86:1067--1070, 2008.

\end{thebibliography}

\end{document}